\RequirePackage{fix-cm}
\documentclass[smallextended]{svjour3}       
\smartqed  

\usepackage{amsgen}
\usepackage{amssymb}
\usepackage{setstack}
\usepackage{bbold}
\usepackage{mathrsfs}
\usepackage{graphicx}

\def\makeheadbox{\relax}
\def\be{\begin{equation}}
\def\ee{\end{equation}}
\def\bea{\begin{eqnarray}}
\def\eea{\end{eqnarray}}
\def\l{\label}
\def\c{\cite}
\def\r{\ref}
\def\p{\phantom}

\renewcommand\ss{\mathscr{S}} 
\newcommand\cc{\mathscr{C}} 
\newcommand{\diag}{\mathop{\rm diag}\nolimits}
\newcommand\ot{\mathcal{O}}
\newcommand\rt{\mathcal{R}}

\newcommand\curl{\mathop{\rm curl}\nolimits}
\newcommand\curlop[1]{\curl\nhalf #1} 
\newcommand\curlH{\curlop{H}}

\renewcommand\div{\mathop{\rm div}\nolimits}

\def\beq{\begin{equation}}
\def\eeq{\end{equation}}

\newcommand\fperp[1]{\half{}^\perp{\nhalf #1}}
\newcommand\fpar[1]{\half{}^\parallel{\nhalf #1}}
\newcommand\fscalar[1]{\half{}^\circ{\nhalf #1}}
\newcommand\fvec[1]{\half{}^\dagger{\nhalf #1}} 
\newcommand\ftensor[1]{\half{}^\ddagger{\nquart #1}}

\newcommand\fscalarE{\fscalar{\nhalf E}}
\newcommand\fvecE{\fvec{\nhalf E}}
\newcommand\ftensorE{\ftensor{\nhalf E}}
\newcommand\fscalarH{\fscalar{\nhalf H}}
\newcommand\fvecH{\fvec{\nhalf H}}
\newcommand\ftensorH{\ftensor{\nhalf H}}
\newcommand\fscalarS{\fscalar{\nhalf S}}
\newcommand\fvecS{\fvec{\nhalf S}}
\newcommand\ftensorS{\ftensor{\nhalf S}}

\newcommand\fscalarsigma{\fscalar{\nquart \sigma}}
\newcommand\fvecsigma{\fvec{\nhalf\sigma}\nquart}
\newcommand\ftensorsigma{\ftensor{\nhalf\sigma}\nquart}

\newcommand\ga{S}
\newcommand\gb{T}
\newcommand{\negminispace}{\kern-.016667em} 

\newcommand{\half}{\kern.083333em}   
\newcommand{\quart}{\kern.0416675em}  
\newcommand{\nhalf}{\kern-.083333em}   
\newcommand{\nquart}{\kern-.0416675em}  

\begin{document}

\title{The Magnetic Part of the Weyl Tensor, and the Expansion of Discrete Universes}


\author{Timothy Clifton         \and \newline
       Daniele Gregoris		\and \newline
       Kjell Rosquist
}


\institute{T. Clifton \at
              School of Physics and Astronomy, Queen Mary University of London, UK. \\
              \email{t.clifton@qmul.ac.uk}           
           \and
           D. Gregoris \at
              Department of Mathematics and Statistics, Dalhousie University, Canada.
              \and
              K. Rosquist \at
              Department of Physics, Stockholm University, 106 91 Stockholm, Sweden. 
}

\date{}

\maketitle

\begin{abstract}
We examine the effect that the magnetic part of the Weyl tensor has on the large-scale expansion of space. This is done within the context of a class of cosmological models that contain regularly arranged discrete masses, rather than a continuous perfect fluid. The natural set of geodesic curves that one should use to consider the cosmological expansion of these models requires the existence of a non-zero magnetic part of the Weyl tensor. We include this object in the evolution equations of these models by performing a Taylor series expansion about a hypersurface where it initially vanishes. At the same cosmological time, measured as a fraction of the age of the universe, we find that the influence of the magnetic part of the Weyl tensor increases as the number of masses in the universe is increased. We also find that the influence of the magnetic part of the Weyl tensor increases with time, relative to the leading-order electric part, so that its contribution to the scale of the universe can reach values of $\sim 1\%$, before the Taylor series approximation starts to break down.
\end{abstract}

\newpage

\section{Introduction}

The large-scale expansion of the Universe is usually taken to be dominated by the Newtonian part of the gravitational field. This idea probably originates from the close correspondence between the Friedmann equations of general relativity and the equations that govern Newtonian cosmologies \cite{milne}, but also has strong support from rigourous constructions that are based on perturbative expansions of Einstein's field equations \cite{viraj,virajX}. Nevertheless, there are good reasons to be interested in relativistic effects in cosmology. These include the apparent existence of Dark Energy, as well as the dawn of the age of precision cosmology. Put bluntly, relativistic effects need to be understood in order to have faith in the cosmological models that we use to interpret observational data.

Geometrically, the electric part of the Weyl tensor is sufficient to determine the Newtonian part of the free gravitational field. The magnetic part of the Weyl tensor, on the other hand, describes other aspects of the relativistic gravitational field \cite{buchert} (both of these objects are defined in Section \r{sec:formalism}, below). Well known relativistic effects such as frame-dragging and gravitational radiation require a non-vanishing magnetic Weyl tensor in order to exist, but the effects of the magnetic part of the Weyl tensor on the large-scale expansion of the Universe are still largely unknown. This is partly due to an absence of realistic cosmological models that have non-zero magnetic Weyl curvature, and that could therefore be considered to be non-silent \cite{bruni}. This situation is, however, changing.

It has recently been demonstrated that cosmological models that contain regularly-arranged discrete masses can generate a non-zero magnetic Weyl tensor, even if no such curvature existed in their initial data \cite{KHB}. The authors of this study considered the effect that this tensor has on the large-scale expansion of a universe that contains eight black holes, using both the leading-order term in a Taylor series expansion, and by numerically integrating the evolution equations of the initially silent space. We extend their study by calculating both the leading-order and next-to-leading-order parts of the relevant series expansion, and by using these results to determine the effect of the magnetic part of the Weyl tensor on the large-scale expansion of all cosmological models that contain regularly arranged discrete masses in a closed cosmology. The phenomenology of such models is interesting, as they can be considered a first approximation to the type of universe within which we actually live. More mathematically, they provide a tractable way to formulate $n$-body cosmology as a relativistic initial value problem. They therefore provide an ideal arena within which to study relativistic gravitational effects in cosmology.

Of course, the magnetic part of the Weyl tensor is a frame-dependent object, and its magnitude will change when different sets of observers are considered. However, in cosmology it is natural to consider sets of time-like curves that are both geodesic, and (in some sense) comoving with the objects that exist within the space-time. It is with respect to  observers following such a set of curves that one can talk about ``cosmological expansion'', and it is with respect to the same set of curves that we will talk about ``the magnetic part of the Weyl tensor''. The influence of the magnetic part of the Weyl tensor on the cosmological expansion is of special interest because virtually all cosmological solutions of Einstein's equations are derived under assumptions that force it to be zero (for cosmologically interesting congruences of curves). Such situations are, however, highly unrealistic, as observers that follow the geodesic set of time-like curves that describe the world-lines of real galaxies will certainly experience the consequences of this part of the curvature of space-time, which is in general non-zero. We must therefore precisely quantify the effects of the magnetic part of the Weyl tensor on the large-scale expansion of space, if we are to fully understand the recessional velocity of galaxies in the real Universe (and all of its associated consequences).

In Section \r{sec:models} we describe in detail the cosmological models that we will be using in this study. They consist of regularly-arranged discrete masses in a closed space, and are formulated as a relativistic initial value problem. In Section \r{sec:formalism} we then introduce the compact formalism that will be used in the rest of the paper. This starts by using a 1+3 decomposition of the geometric and kinematic quantities involved, and finishes with a 1+1+2 decomposition that can be efficiently used to exploit the symmetries of the problem. The explicit form of the symmetry restrictions are then determined in Section \r{sec:reflection}, where it is shown that the influence of the magnetic part of the Weyl tensor on the relevant evolution equations does not necessarily vanish. Section \r{calcs} then contains some lengthy calculations to determine the coefficients of a Taylor series expansion that can be used to include the effect of the magnetic Weyl tensor on the large-scale expansion of space. In Section \r{sec:results} we present the numerical results for each of the lattice universes that we will be considering, before concluding in Section \r{sec:discussion}. 

Throughout the paper we use the first half of the Latin alphabet to denote space-time coordinate indices, and the second half to denote spatial coordinate indices. Greek letters are reserved to denote spatial frame indices, where they are required.

\section{Lattice Models of the Universe}
\l{sec:models}

The space-times we wish to consider are those that have come to be known as ``lattice models'', in some parts of the cosmology literature. These models have a periodic structure, and are constructed from a number of regularly arranged cells, each of which are identical to one another. Such models have been studied in a number of different contexts over the past few years, as they offer a simple enough setting to provide concrete answers to questions of direct physical interest. 

The particular type of lattices we wish to consider are those in which the topology of the cosmological region is a hypersphere. Furthermore, we want to consider situations where the only mass present is in the form of black holes, without angular momentum or electric charge. We position one single black hole at the centre of every lattice cell, so that the cosmological region as a whole contains a regular array of these objects. The geometry of the space-time can then be treated as a vacuum solution of Einstein's equations, with its scale and expansion being prescribed accordingly.

In the rest of this section we will describe the specifics of the six different lattice configurations that we will be studying, as well as the solutions to the constraint equations, and the existence of curves within the space-time that display special rotational symmetries. This recaps results from previous work \cite{Clifton_etal:2012,Clifton_etal:2013,Clifton_etal:2014}.

\subsection{The Six Closed Lattices}

Before solving for the geometry of space-time, we need to understand how to divide a hyperspherical universe into $n$ identical cells. If we choose each cell to be one of the five possible Platonic solids ({\it i.e.} one of the five possible regular convex polyhedra), then there are six different ways that we can arrange these cells to form a closed lattice. Each of these configurations corresponds directly to one of the six regular, convex polychora that exist in four dimensions. They are \cite{coxeter}:
\begin{itemize}
\item {\bf 5-cell.} This lattice consists of five tetrahedral cells. Three cells meet along every cell edge, and there are a total of five vertices (the points where cell corners meet). This is the smallest lattice that can be constructed from regular convex polyhedra.
\item {\bf 8-cell.} This structure is composed of eight cubes, arranged so that three cells meet along every cell edge, giving a total of sixteen vertices. It corresponds to the space-time studied numerically in \cite{KHB,8cell}.
\item {\bf 16-cell.} This lattice is made from sixteen tetrahedral cells, with four cells meeting at every cell edge. It is dual to the 8-cell (meaning that interchanging vertices and cell centres gives an 8-cell lattice).
\item {\bf 24-cell.} The 24-cell is constructed from twenty four octahedral cells. Three of these cells meet along every cell edge, and there are a total of twenty four vertices. The 24-cell is self-dual, as it has the same number of cells and vertices.
\item {\bf 120-cell.} This structure contains one hundred and twenty dodecahedral cells, with three cells meeting at every edge. There are six hundred vertices in this structure, meaning that it is dual to the 600-cell, described below.
\item {\bf 600-cell.} This is the largest of all lattices that fits our specifications. It is built from six hundred tetrahedral cells, with five of these meeting along every edge.
\end{itemize}
In the rest of this section we will outline how Einstein's constraint equations can be solved for these lattices, as well as how curves that exhibit local rotational symmetry can be identified.

\subsection{Initial Data}

If we treat general relativity as an initial value problem, then we can split Einstein's equations into constraint and evolution equations. The former of these can then be written as a Hamiltonian constraint and a momentum contraint, such that, in vacuum,
\bea
{}^{(3)}R+K^2 - K_{ij} K^{ij} = 0 \l{con1} \\
D_j K_i^{\p{i}j}- D_i K = 0 \l{con2} \, ,
\eea
where ${}^{(3)}R$ is the Ricci scalar of the geometry on the initial hypersurface, $\Sigma$, and $K_{ij}$ is its extrinsic curvature in the embedding space-time. The $D_i$ derivative is used here to denote a covariant derivative within $\Sigma$, and $K$ is the trace of $K_{ij}$.

Eqs.\ (\r{con1}) and (\r{con2}) are, in general, very difficult to solve. However, we can simplify the problem dramatically if we choose our initial hypersurface carefully. For example, if we choose it to be symmetric under time reversal then we automatically have $K_{ij}=0$. The Hamiltonian constraint then reduces to
\be
{}^{(3)}R =0 \, , \l{con3}
\ee
while the momentum constraint is identically satisfied. This equation is much easier to solve, and is the basis of the study of geometrostatics (the gravitational analogue of the field of electrostatics \cite{misner}). We will assume initial data of exactly this type in what follows, and find solutions to the constraint equation (\r{con3}) that describe the maximum of expansion of our lattices. These results were presented for the first time in \c{Clifton_etal:2012}.

A metric ansatz for the initial geometry can be chosen as follows:
\be
dl^2 = \psi^4 \left( d\chi^2 + \sin^2 \chi d \Omega^2 \right) \, , \l{sol1}
\ee
where $d\Omega^2 = d\theta^2 + \sin^2 \theta d \phi^2$ is the line-element of a 2-sphere, and where $\psi=\psi (\chi, \theta ,\phi)$ acts as the square root of the scale factor. The time-symmetric Hamiltonian constraint then reduces to
\be
{\bar{\nabla}}^2 \psi = \frac{3}{4} \psi \, , \l{con4}
\ee
where $\bar{\nabla}^2$ is the Laplacian operator on the conformal $3$-sphere. This equation is now linear in $\psi$, meaning that if a solution can be found for a single black hole, then we can simply super-impose this with other solutions of the same form to find multi-black hole solutions. This is exactly what we need to form our lattice models of the Universe.

A simple solution to Eq.\ (\ref{con4}) is given by $\psi = \sqrt{\tilde{m}}/(2 \sin \left(\frac{\chi}{2}\right))$, where $\tilde{m}$ is a constant. This function has a pole at $\chi=0$, and looks very much like the gravitational field one might expect for a point-like particle located at that position. Of course, there is nothing special about the point $\chi=0$, and we can generate any number of similar functions by rotating the coordinate system so that $\{\chi,\theta,\phi\} \rightarrow \{\chi_i,\theta_i,\phi_i\}$, and then replacing $\chi$ by $\chi_i$ in the denominator. This gives the solution for $N$ such particles as
\be
\psi = \sum_{i=1}^N \frac{\sqrt{\tilde{m}_i}}{2 \sin \left( \frac{\chi_i}{2} \right)} \, , \l{sol2}
\ee
where $\chi_i$ should be understood as the value of $\chi$ in a coordinate system that has been rotated so that the $i$th particle appears at $\chi_i=0$. This is just what we need to construct our lattice cosmologies, as it is a solution that corresponds to $N$ arbitrarily located black holes, in a universe that is at its maximum of expansion.

In the sections that follow, we will use the solution given by Eqs.\ (\r{sol1}) and (\r{sol2}) as initial data for our cosmology. This initial data is exact, but has some subtleties to its interpetation. The parameter $\tilde{m}_i$ that appears in Eq.\ (\ref{sol2}) looks like it corresponds to the mass of a particle positioned at $\chi_i=0$, but it is not the proper mass of the $i$th black hole. This must be found by looking at the geometry of the space in the limit $\chi_i \rightarrow 0$, and comparing this to a time-symmetric slice through the Schwarzschild solution. One then finds that the proper mass of each of the black holes is a function of the location and $\tilde{m}_i$ of each of the other black holes. The parameter $\tilde{m}_i$ can then be interpreted as an ``effective mass'', which includes information about the gravitational field of the gravitational potential energy, as well as that from the proper mass of each of the individual black holes \c{Clifton_etal:2012,mass}. As we want all of the black holes in our lattice models to have the same mass, we choose $m_i=m$ for all values of $i$.

\subsection{Curves with a Rotationally Symmetric Tangent Space}

There exists a number of curves, within our lattices, that display Local Rotational Symmetry (LRS) \c{Clifton_etal:2013,Clifton_etal:2014}. Perhaps the simplest of these curves to visualize is the example of a cell edge. In each of the lattice structures listed above there were between three and five identical cells meeting along every cell edge. This means that the geometry of space-time in the vicinity of these curves must be invariant under a rotation that interchanges the cells. If we pick a set of orthonormal frame vectors for our 3-space $\{{\bf e}_1,{\bf e}_2,{\bf e}_3 \}$, and align ${\bf e}_1$ with this special curve, then the rotational symmetry implies \c{Clifton_etal:2013}
\be
V_2=V_3 =0 =T_{23} \qquad {\rm and} \qquad T_{22}=T_{33} \, , \l{lrs}
\ee 
where $V_\alpha$ and $T_{\alpha \beta}$ are the spatial frame components of any vector ${\bf V}$, and any symmetric tensor ${\bf T}$. Furthermore, the commutators of the basis vectors are restricted by \c{Clifton_etal:2013}
\be
\gamma^{1}_{\p{1}21} = \gamma^{1}_{\p{1}31} =0, \qquad \gamma^{2}_{\p{2}12} =\gamma^{3}_{\p{3}13} \qquad {\rm and} \qquad \gamma^{3}_{\p{3}12}= \gamma^{2}_{\p{2}13} \, , \l{lrs2}
\ee
where $[{\bf e}_{\alpha},{\bf e}_{\beta} ] = \gamma^{\epsilon}_{\p{\epsilon} \alpha \beta} {\bf e}_{\epsilon}$. The results in Eqs.\ (\r{lrs}) and (\r{lrs2}) are only valid in the tangent spaces of the points that make up these LRS curves, but they are significant restrictions.

As well as cell edges, there are other curves in our lattices that also exhibit rotational symmetry \c{Clifton_etal:2014}. They are given by the lines that connect the horizon of the black hole at the cell centre with the cell corner, and the line that connects this horizon with the centre of a cell face. Such curves are illustrated in Fig.\ \r{cubefig}, for the particular case of a cubic cell. They will be the curves that we solve for in Section \r{sec:results}. The curves that connect the centre of the cell with the cell corner, and the centre of a cell face, are formally infinitely long in the initial data, which is why we have chosen to cut them off at the black hole horizon. They then have finite proper length.

The location of the horizon will be determined by assuming the horizon to be initially non-expanding. In this case, the location of the horizon along LRS curves is found by looking for solutions to the following equation \c{Clifton_etal:2013}:
\be
D_{{\bf e}_1} E_{11} =0 \, ,
\ee
where ${\bf e}_1$ is again aligned with the direction of the LRS curves, and $E_{11}$ is the frame component of the electric part of the Weyl tensor. The location of the horizon at subsequent times is found by propagating null geodesics along the LRS curves, from the initial location (as can be seen to be required by the null Raychaudhuri equation \c{Clifton_etal:2013}).

\begin{figure}[t!]
\begin{centering}
\includegraphics[width=4in]{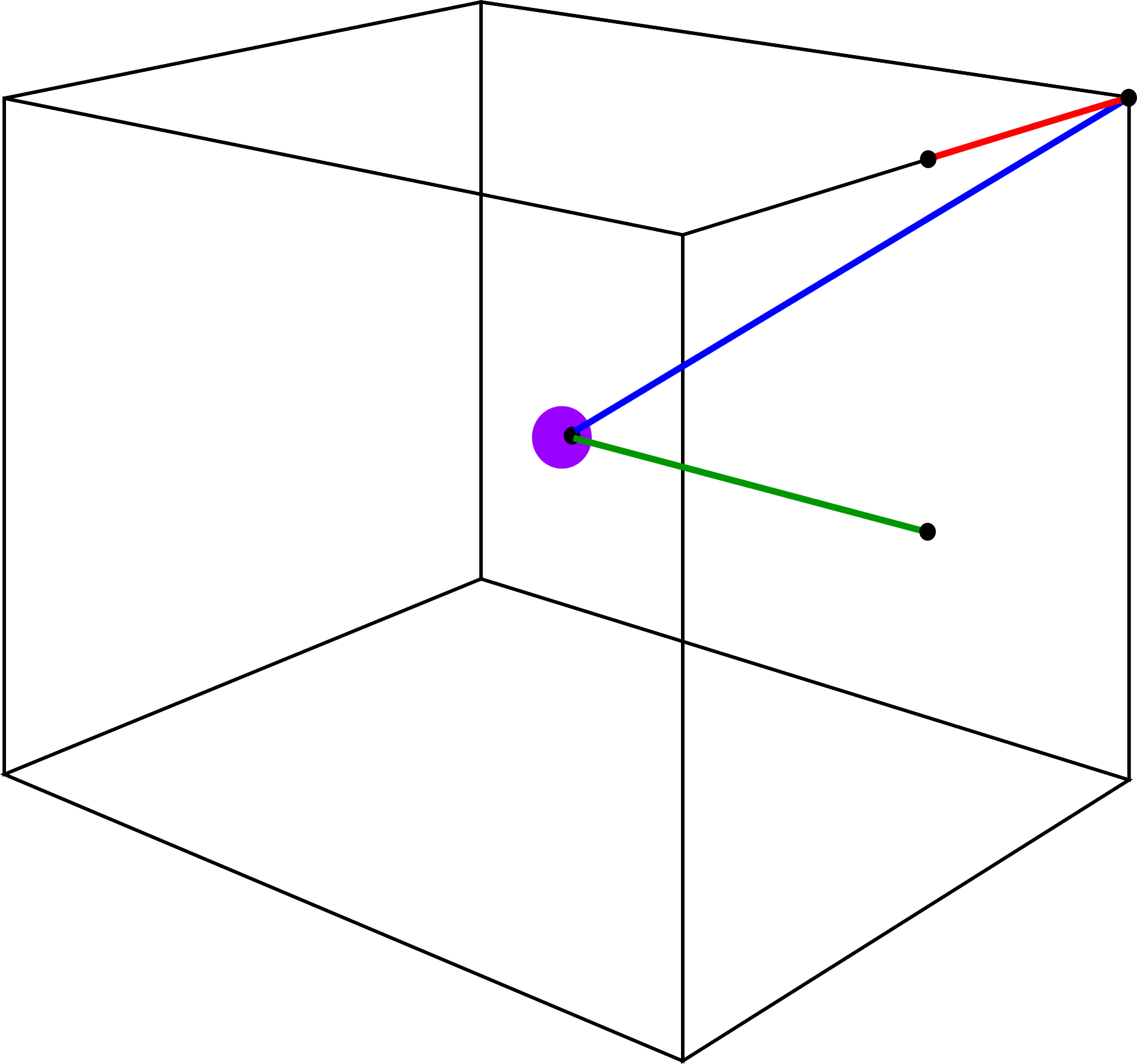}
\par\end{centering}
\caption{An illustration of the curves that connect the centre of the cell with the centre of a cell face (green), and the centre of a cell with a cell corner (blue). Also displayed is a curve that connects the centre of a cell edge with a cell corner (red). These are the three curves whose evolution will be solved for in Section \r{sec:results}.}
\centering{}\label{cubefig}
\end{figure}

\section{Covariant Formalism and Field Equations}
\l{sec:formalism}

In order to exploit the symmetries of a space-time it is often convenient to decompose the geometric objects that exist within it. This can be achieved by picking out vector fields that are either aligned with, or orthogonal to, certain invariantly defined sub-spaces, and then by defining projection operators associated with these vectors. Within the field of cosmology, the most widely applied decomposition of this type is the $1+3$-decomposition pioneered by Ehlers \c{ehlers}, and developed by Ellis and others (see {\it e.g.} \cite{ellis1,ellis2}). If there also exists a preferred space-like direction within the space, as in LRS space-times \cite{LRS}, then the $1+1+2$ decomposition of Clarkson becomes a useful tool \cite{clarkson}. Going further, and decomposing using a full set of four mutually orthogonal unit vectors, one is then led to the orthonormal frame approach \cite{frames}. In this section we will outline the $1+3$ and $1+1+2$ approaches, as relevant for our study.

\subsection{The $1+3$ Decomposition}

If there exists a time-like unit vector field, {\bf u}, then we can define tensors that project parallel and othogonal to this vector, respectively, as
\be
\l{projection}
U^a_{\p{a} b} := - u^a u_b \qquad {\rm and } \qquad h^a_{\p{a} b} := \delta^a_{\p{a} b} + u^a u_b \, ,
\ee
where $\delta^a_{\p{a} b}$ indicates the Kronecker delta. We can also project all derivatives along and orthogonal to $u^a$, respectively, by defining
\bea
\dot{S}^{a \dots b}_{\p{a \dots b} c \dots d} &:=& u^e \nabla_e S^{a \dots b}_{\p{a \dots b} c \dots d}\\
D_a S^{b \dots c}_{\p{b \dots c} d \dots e} &:=& h_a^{\p{a} f} h^b_{\p{b} g} \dots h^c_{\p{c} h} h_d^{\p{d} i} \dots h_e^{\p{e} j} \nabla_f S^{g \dots h}_{\p{g \dots h} i \dots j} \, ,
\eea
where $S^{a \dots b}_{\p{a \dots b} c \dots d}$ is any tensor. It can immediately be seen that $\dot{h}_{ab}= 2 \dot{u}_{(a} u_{b)}$ and $D_a h_{bc}=0$, which can be considered the evolution and constraint equations for $h_{ab}$.

The covariant derivative of $u_a$ can then be decomposed into its irreducible components, such that
\be
\label{1+3}
\nabla_a u_b = -u_a \dot{u}_b + \frac{1}{3} \Theta h_{ab} + \sigma_{ab} + \omega_{ab} \, ,
\ee
where $\dot{u}^a= u^b \nabla_b u^a$ is the acceleration vector, $\Theta := u^{a}_{\p{a};a}$ is the expansion scalar, $\sigma_{a b} := \dot{u}_{(a} u_{b)} + \nabla_{(a} u_{b)} - \frac{1}{3} \Theta h_{ab}$ is the symmetric trace-free shear tensor, and $\omega_{ab} := u_{[a} \dot{u}_{b]} + \nabla_{[a} u_{b]}$ is the skew-symmetric vorticity tensor. These last three quantities describe the rate at which the integral curves of $u^{a}$ expand, squash and twist around each other. It is straightforward to verify that $U^{a b} \sigma_{bc} = U^{a b} \omega_{bc} =0 = U^{ab} \dot{u}_b$, meaning that all of these quantities are already projected orthogonal to $u^a$.

The fundamental object of interest for describing the free gravitational field, and which is required to complete a description of the vacuum scenario, is the Weyl tensor, $C_{abcd}$. The existence of the time-like vector field {\bf u} can be used to split the Weyl tensor into ``electric'' and ``magnetic'' parts, defined respectively by
\be
E_{ab} := C_{acbd} u^c u^d \qquad {\rm and} \qquad H_{ab} := \frac{1}{2} \epsilon_{acd} C^{cd}_{\p{cd}be} u^e \, ,
\ee
where $\epsilon_{a b c}=\eta_{abcd}u^d$ is the spatial skew symmetric permutation tensor (defined such that $\eta_{0123} = -\sqrt{\vert g \vert }$). The electric part of the Weyl tensor contains all information about the tidal forces due to gravity. The magnetic part, on the other hand, contains all other information about the Weyl curvature. We also have $\dot{\epsilon}_{abc} = \eta_{abcd} \dot{u}^d$ and $D_a \epsilon_{abc}=0$, which can be thought of as the evolution and constraint equations for $\epsilon_{abc}$.

\subsection{The Vacuum Field Equations}

In order to write the field equations in the most concise form possible, it is convenient to introduce the definitions of the covariant divergence and curl, which when operating on rank-2 tensors are given by
\be
(\div \ga)_a := D^b \ga_{ab} \qquad {\rm and} \qquad  \curl \!  \ga_{ab}
                  := \epsilon_{cd(a} D^c \ga_{b)}{}^d \, .
\ee
It is also convenient to introduce the projected, symmetric, trace-free part of a rank-2 tensor by using angular brackets, such that
\bea 
\ga_{\langle ab\rangle} := h_a{}^c h_b{}^d \ga_{(cd)} -\frac{1}{3} \ga_{cd}h^{cd} h_{ab} \, .
\eea
Note that this means $\sigma_{\langle ab \rangle}= \sigma_{ab}$, $E_{\langle ab \rangle}= E_{ab}$, and $H_{\langle ab \rangle}= H_{ab}$, as these tensors are already projected, symmetric and tracefree.

With the definitions of all of these quantities and operators in hand, it is now possible to write the vacuum evolution equations for a geodesic ($\dot{u}^a=0$), irrotational ($\omega_{ab}=0$) flow as \cite{Maartens:1997}
\bea\label{eeq}
\dot{\Theta} &=& -\frac{1}{3}\Theta^2 - \sigma_{ab}\sigma^{ab}\\
\dot{\sigma}_{ab} &=& -\frac{2}{3} \Theta \sigma_{ab} -\langle\sigma,\sigma\rangle_{ab} - E_{ab} \l{evs}\\
\dot{E}_{ab} &=& -\Theta E_{ab} + 3\langle\sigma,E\rangle_{ab} +\curl \!H_{ab} \label{evE}\\
\dot{H}_{ab} &=& -\Theta H_{ab} + 3\langle\sigma,H\rangle_{ab} -\curl \!E_{ab} \, , \label{evH} 
\eea
with the constraint equations taking the form
\bea
(\div\sigma)_a &=& \frac{2}{3} D_a \Theta \\
\curl\sigma_{ab} &=& H_{ab} \label{c4}\\
(\div E)_a &=&  [\sigma,H]_a \l{c5}\\
(\div H)_a &=& -[\sigma,E]_a \, . \label{eeqlast}
\eea
The bracket operators in these equations are defined as $[\ga,\gb]_a := \epsilon_{abc} \ga^b{}_d \gb^{cd}$, and $\langle \ga, \gb\rangle_{ab} := \ga_{c\langle a} \gb_{b\rangle}{}^c$. This extremely compact way of writing the field equations will greatly aid the manipulations we perform in Section \ref{calcs}. It also has the distinct advantage that all quantities involved have a direct physical interpretation, allowing for a better intuitive understanding of the physical set-up.

\subsection{The $1+1+2$ Decomposition}

Just as in the previous sections, and following Clarkson \cite{clarkson}, if there exists a space-like unit vector field, ${\bf n}$, then we can decompose all geometric objects into parts that are parallel and orthogonal to this field.  The goal here is to first perform the decomposition with respect to $u^a$, and then to take the tensors projected into the space orthogonal to $u^a$ and decompose them further with respect to $n^a$. To this end, we introduce the following projection tensors:
\be
N^{a}_{\p{a}b} := n^a n_ b \qquad {\rm and} \qquad f^{a}_{\p{a}b} := h^a_{\p{a} b} - n^a n_b \, ,
\ee
where $h_{ab}$ is the projection tensor from above. The tensor $N^a_{\p{a} b}$ can operate on any geometric object, and projects it along $n^a$. The tensor $f^a_{\p{a}b}$, on the other hand, only makes sense as a projection tensor when it operates on objects that have already been projected using $h^a_{\p{a}b}$. If $n^a$ is surface forming, then $f^{a}_{\p{a}b}$ projects such objects onto the 2-dimensional ``sheets'' orthogonal to $n^a$.

Under these definitions, we can irreducibly decompose any spatially projected vector, $V^a=h^a_{\p{a}b} V^b$, into a scalar part $\fpar{V}:= V^a n_a$, and a transverse vector part $\fperp{V_a} := f_a{}^{b}\,V_b$. The projected vector can then be written as
\be
V_a = \fpar{V_a} + \fperp{V_a}
\ee
where $\fpar{V_a} =  N_a^{\p{a}b} V_b$. These definitions mean that $f^a_{\p{a}b} \fpar{V_a} = N^a_{\p{a}b} \fperp{V_a}=0$, which shows that $\fpar{V_a}$ has no parts in the sheet, and $\fperp{V_a}$ has no part parallel to $n^a$.

The other geometric objects we wish to decompose with respect to $n^a$ are the symmetric trace-free tensors defined above. Any such tensor can be irreducibly decomposed into a scalar $\fscalar{S} := N^{ab} S_{ab}$, a transverse vector $\fvec{S}_a := f_a{}^b n^c S_{bc}$, and a transverse and tracefree tensor $\ftensor{S}_{ab} := ( f_{(a}{}^c f_{b)}{}^d -\frac{1}{2}f_{ab}f^{cd} ) S_{cd} $. The full symmetric and trace-free tensor can then be written in terms of these fundamental objects as
\begin{equation}
S_{ab} = \fscalar{S}_{ab} + \fvec{S}_{ab} + \ftensor{S}_{ab} \, ,
\end{equation}
where $\fscalar{S}_{ab} = \fscalar{S} ( N_{ab} - \frac{1}{2} f_{ab})$, $\fvec{S}_{ab} = 2 n_{(a} \nhalf\fvec{S}_{b)}$, and $\ftensor{S}_{ab} = ( f_{(a}{}^c f_{b)}{}^d -\frac{1}{2}f_{ab}f^{cd} ) S_{cd}$. These three objects are each symmetric, and trace-free with respect to $h_{ab}$. They are the only three such objects that can be constructed from a single power of $\fscalar{S}$, $\fvec{S}_a$ and $\ftensor{S}_{ab}$, and as such they specify a unique decomposition of $S_{ab}$. In the next section we will consider the restrictions that reflection symmetry imposes upon these objects, and why this means that $H_{ab}$ can have an effect on the large-scale expansion of space.

\section{Reflection Symmetric Surfaces}
\l{sec:reflection}

The presence of reflection symmetric surfaces can greatly simplify the formulation of problems in general relativity. Indeed, it is the presence of a reflection symmetry in the time-like direction that allowed the constraint equations to be solved for a lattice universe \cite{Clifton_etal:2012}. Reflection symmetry in the space-like directions normal to certain $2+1$-dimensional surfaces within the space-time also lead to the notion of ``piecewise silence'', as gravitational waves are forbidden from travelling between lattice cells, or the various chambers within cells \cite{Clifton_etal:2014}. In this section we will investigate the properties of reflection symmetric surfaces further, by formulating their consequences for geometric objects in the $1+1+2$ decomposition, and by considering the situation where multiple reflection symmetric surfaces intersect (as often occurs in our lattice universe construction).

\subsection{Polar Tensors and Axial Tensors}

When determining the consequences of reflection symmetry for tensorial quantities it is important to distinguish between polar and axial tensors. If a tensorial object can be defined without reference to a specific frame, then that object is said to be a ``polar tensor''. By contrast, if such an object is defined with respect to some fixed direction or combination of directions, then that object is said to be an ``axial tensor''. 

The 3-dimensional Levi-Civita tensor, $\epsilon_{abc}$, is the archetypical axial tensor. If $\{{\bf e_1},{\bf e_2}, {\bf e_3}\}$ constitute a set of orthonormal spatial frame vectors, then the frame components of this tensor are given by $\epsilon_{123} =1$, and all permutations. In other words, the Levi-Civita tensor defines a preferred orientation of the spatial frame. A reflection, or any other improper orthonormal transformation, changes the orientation of the spatial frame. Reflections therefore change the sign of $\epsilon_{abc}$, meaning that it should be properly identified as a pseudotensor, rather than as a tensor. This distinction is extremely important when determining the consequences of reflection symmetry on geometric quantities.

To illustrate this in a formal way, we note that an inversion can be represented by the orthonormal transformation matrix $I = -\mathbb{1}$, where $\mathbb{1}$ denotes the 3-dimensional unit matrix. Applying the standard tensor transformation formula for an orthogonal transformation matrix $\ot^a{}_b$ would then give
\begin{equation}\label{ttf}
\ot (S_{abc \dots}) = \ot^d{}_a \, \ot^e{}_b \, \ot^f{}_c  \dots\, S_{def\dots} \, .
\end{equation}
If we take $S_{abc \dots} = \epsilon_{abc}$ and $\ot=I$ then this law would give us $\ot(\epsilon_{abc}) = -\epsilon_{abc}$. The sign change indicates that the transformed object, $\ot (\epsilon_{abc})$, has opposite parity to the untransformed object, $\epsilon_{abc}$. This is an unwanted feature if we require the parity of the Levi-Civita tensor (right or left-handedness) to be invariant under any orthogonal transformation. The standard way around this is to modify the transformation rule for pseudotensors, so that it becomes \cite{Miller:1972}
\begin{equation}\label{ptf}
\ot (S_{abc}\dots) = \det(\ot) \ot^d{}_a \, \ot^e{}_b \, \ot^f{}_c \dots\, S_{def\dots} \, .
\end{equation}
We can then transform tensors according to Eq.\ (\r{ttf}), whilst transforming pseudotensors according to the expression in Eq.\ (\r{ptf}). The extra factor of $\det(\ot)$ in this latter equation means that any quantity that contains an odd number of Levi-Civita tensors picks up an extra minus sign under any improper orthogonal transformation, such as a reflection.

\subsection{Implementing reflection symmetries}

Consider a reflection symmetry in a spatial hypersurface. Such a symmetry has fixed points that form an invariant symmetry surface, $\ss$. Let $\{{\bf n},{\bf k}, {\bf l}\}$ be an orthonormal frame adapted to the symmetry in such a way that ${\bf n}$ is orthogonal to $\ss$, and consequently that ${\bf k}$ and ${\bf l}$ are parallel to $\ss$. The reflection symmetry in the tangent space of a point in $\ss$ can then be represented by the orthonormal frame transformation matrix $\rt^a_{\p{a}b} = \diag(-1,1,1)$ for a polar tensor, and by $\tilde\rt = -\rt$ for an axial tensor. The restrictions on a tensor ${\bf Q}$, imposed by invariance under a reflection symmetry, is then determined by the equations
\begin{equation}\label{symcond}
\rt {\bf Q} = {\bf Q} \qquad {\rm or} \qquad \tilde\rt {\bf Q} = {\bf Q} \, ,
\end{equation}
depending on whether ${\bf Q}$ is a polar or an axial tensor, respectively. We note that, for the specific fields that appear in Eqs.\ (\r{eeq})-(\r{eeqlast}), the set $\{ \Theta, \sigma_{ab}, E_{ab} \}$ are polar, while $H_{ab}$ is axial (as it involves the Levi-Civita pseudotensor). The only operation in the field equations that changes the parity of an object is the $\curl$, which means that $\curl\nhalf H_{ab}$ is polar, while $\curl\nhalf\sigma_{ab}$ and $\curl\nhalf  E_{ab}$ are axial tensors. Decomposed tensor parts, such as $\fvecE_{ab}$, always have the same parity as the full tensor from which they are derived. 

To calculate the action of the reflection $\rt$, let us first consider the case of scalars. We note that the only change possible for a scalar is the change of sign that affects pseudoscalars under an improper transformation, according to the pseudoscalar version of Eq.\ (\ref{ptf}). This implies that all pseudoscalars must vanish on $\ss$, leading to the restrictions $\fscalar{(\curl\sigma)} =\fscalar{(\curl\nhalf  E)} =\fscalarH =0$. To see the effect on vectorial and tensorial objects we can express the transformation matrix in the form $\rt^a_{\p{a}b} = f^a_{\p{a} b} - N^a{}_b$. The action on the vector part of a tensor, $S_{ab}$, is then given by%
\footnote{As a technical note, it can be remarked that $n^a$, when regarded as a fixed vector, will be an axial vector. This has the effect that $\fvecS_a$ and $\fvecS_{ab}$ have opposite parity, as their definitions include one and two factors of $n^a$, respectively. This issue is irrelevant for the scalars $\fscalarS$ and $\fscalarS_{ab}$ as both contain an even number of factors of $n^a$ (2 and 0, respectively).}
\begin{equation}
\rt (\fvecS_{ab}) = \rt^c{}_a \rt^d{}_b \fvecS_{cd}
    = 2(f^c{}_a- N^c{}_a)(f^d{}_b-N^d{}_b) N_{(c}^{\p{(c} e} f_{d)}{}^f S_{ef}
    = -\fvecS_{ab} \, .
\end{equation}  
Therefore, in order to fulfil the symmetry condition given in Eq.\ (\ref{symcond}), the vectorial part, $\fvecS_{a}$, of a polar tensor, $S_a$, must vanish on $\ss$. It then follows that $\fvecsigma_a =\fvecE_a =\fvec(\curl\nhalf  H)_a =0$ on $\ss$. Applying the reflection transformation in an analogous way to the tensor part of $S_{ab}$ gives $\rt(\ftensorS_{ab}) = \ftensorS_{ab}$. Hence, the polar tensor part is not affected by the reflection, and is therefore unrestricted. The axial tensor part, on the other hand, is required to vanish, due to the determinant appearing in Eq.\ (\ref{ptf}). This implies that $\ftensor{(\curl\nhalf  E)_{ab}} =\ftensorH_{ab} =0$. 

In summary, the tensor parts remaining on a reflection-symmetric surface, after the symmetry restrictions imposed by Eq.\ (\r{symcond}), are $\Theta$, $\fscalarsigma$, $\ftensorsigma_{ab}$, $\fvec(\!\curl\nhalf \sigma)_a$,  $\fscalarE, \ftensorE_{ab}$, $\fvec(\!\curl E)_a$, $\fvecH_a$, $\fscalar(\!\curl\nhalf H)$, and $\ftensor(\!\curl\nhalf H)_{ab}$.

\subsection{Intersection of Two Symmetry Surfaces}

We now consider the symmetry restrictions imposed at points where two reflection invariant surfaces meet. Let us call these two surfaces $\ss_k$ and $\ss_l$, where the subscripts $k$ and $l$ indicate their normals. The intersection of these two surfaces forms a curve, $\cc$, and reflection symmetry about each surface ensures that they meet at right angles to each other. In this case we can choose to adapt the frame in such a way that ${\bf n}$ is parallel to $\cc$. The quantities $\fscalar{(\curl\sigma)}$, $\fscalar{(\curl\nhalf  E)}$ and  $\fscalarH$ now have a different interpretation, but they are again pseudoscalars. We therefore have that
$\fscalar{(\curl\sigma)} = \fscalar{(\curl\nhalf  E)} = \fscalarH =0$.
 
To analyze the restrictions on vectorial and tensorial objects, we note that the reflection transformations associated with $k$ and $l$ can be expressed in the forms
\begin{equation}
   \rt_k: \gamma^a{}_b = h^a{}_b -2 k^a k_b \ ,\qquad 
   \rt_l: \tilde \gamma^a{}_b = h^a{}_b -2 l^a l_b \, .
\end{equation}
Applying $\rt_k$ to the vectorial part of the $1+1+2$ decomposition gives
\begin{equation}
   \rt_k(\fvecS_{ab}) = 2\gamma^c{}_a \gamma^d{}_b n_{(c} f_{d)}^e n^f S_{ef}
                      = \fvecS_{ab} - 4n_{(a}k_{b)}\half k_c \nhalf\fvecS^c \, .
\end{equation}
Applying the symmetry condition from Eq.\ (\ref{symcond}), for a polar tensor, then leads to $k_a \nhalf \fvecS^a = 0$ on $\ss_k$. A similar argument for the reflection symmetry $\rt_l$ gives $l_a \nhalf \fvecS^a = 0$ on $\ss_l$. It follows that $\fvecS^a =0$ on $\cc$, as both of its components must vanish. This implies the restrictions $\fvecsigma_a = \fvecE_a = \fvec{(\curl\nhalf  H)}_a =0$.

For the vector part of an axial vector we get
\begin{equation}
   \tilde\rt_k(\fvecS_{ab}) = -\fvecS_{ab} + 4n_{(a}k_{b)}\half k_c \nhalf\fvecS^c\, .
\end{equation}
In this case, the symmetry condition from Eq.\ (\r{symcond}) takes the traceless symmetric form
\begin{equation}
\l{cab}
   C_{ab} := n_{(a}\fvecS_{b)} - n_{(a} k_{b)}\half k_c \nhalf\fvecS^c = 0 \, .
\end{equation} 
Only the vectorial component of $C_{ab}$ is non-zero, which means the symmetry condition in Eq.\ (\r{cab}) then reduces to
\begin{equation}
   \fvec{C_a} = \frac{1}{2} (\fvecS_a - k_a \half k_b \nhalf\fvecS^b) = 0 \, .
\end{equation}
This tells us that the component of $\fvecS_a$ in the \mbox{$l$-direction} vanishes on $\ss_k$. Similarly, the symmetry $\rt_l$ implies that the component of $\fvecS_a$ in the $k$-direction vanishes on $\ss_l$, and consequently that the vectorial parts of all axial tensors vanish on $\ss$. This gives the restrictions $\fvec(\curl \sigma)_a = \fvec(\curl\nhalf  E)_a = \fvecH_a =0$.   

Let us now consider the consequences of the symmetry conditions, from Eq.\ (\r{symcond}), on the tensorial part of a polar tensor. We find that the reflection operation gives
\begin{equation}
\l{symten}
   \rt_k(\ftensorS_{ab}) = \ftensorS_{ab} - 4 (\ftensorS_{cd}\half k^c l^d) \quart k_{(a} l_{b)} \, ,
\end{equation}
where we have used the relation $\ftensorS_{ab}\half k^b = (\ftensorS_{bc}\half k^b k^c) k_a + (\ftensorS_{bc}\half k^b l^c) \half l_a$.
%
%
The symmetry condition in Eq.\ (\ref{symcond}) then implies $\ftensorS_{ab}\half k^a l^b=0$.
The tensorial parts $\ftensorsigma_{ab}$, $\ftensorE_{ab}$ and $\ftensor(\curl\nhalf  H)_{ab}$ are therefore diagonal, with $\ftensorS_{a b}\half k^a k^b$ 
as their only independent component. As the right-hand side of Eq.\ (\ref{symten}) is symmetric in $k$ and $l$, there are no further restrictions implied by the symmetry action of $\rt_l$. 

Finally, we consider the action of the reflection symmetries on axial tensors. They give
\begin{equation}
   \tilde\rt_k(\ftensorS_{ab}) = \tilde\rt_l(\ftensorS_{ab}) = -\ftensorS_{ab}
                          +4 (\ftensorS_{cd}\half k^c l^d) \quart k_{(a} l_{b)} \, ,
\end{equation}
so that the symmetry conditions in Eq.\ (\ref{symcond}) then become
\begin{equation}
   \ftensorS_{ab} = 2 (\ftensorS_{cd}\half k^c l^d )\quart k_{(a} l_{b)} \, .
\end{equation}
This is the condition that $\ftensorS_{ab}\half k^a l^b$ is the only nonvanishing independent component of $\ftensorS_{ab}$. This restriction applies to $\ftensor(\curl \sigma)_{ab}$, $\ftensorH_{ab}$ and $\ftensor(\curl\nhalf  E)_{ab}$.

In summary, at the intersection of two reflection symmetric surfaces we have that the only non-vanishing tensor parts are given by $\Theta$, $\fscalarsigma$, $\fscalarE$ and $\fscalar(\!\curl\nhalf H)$, the single independent component of the diagonals of $\ftensorsigma_{ab}$, $\ftensorE_{ab}$ and $\ftensor(\!\curl\nhalf H)_{ab}$, and the single independent off-diagonal components of $\ftensor(\!\curl\nhalf \sigma)_{ab}$, $\ftensor(\!\curl\nhalf E)_{ab}$ and $\ftensorH_{ab}$ \footnote{This corrects statements made in \c{Clifton_etal:2013}.}.

\subsection{Intersection of Three or More Symmetry Surfaces}

When three or more reflection symmetric surfaces meet the restrictions on geometric objects are much stricter. Such situations arise along LRS curves when the symmetry along the intersecting surfaces is threefold or higher, as was studied in some detail in \cite{Clifton_etal:2013}. Here we adapt our frame to the symmetry curve in the same manner as in the twofold symmetry case, considered above. As we know that all the vectorial and tensorial parts of every geometric object must vanish in this case \cite{Clifton_etal:2013}, we only need to consider the scalars. As before, the axial scalars must once again vanish, implying that $\fscalar{(\curl \sigma)} = \fscalar{(\curl\nhalf  E)} = \fscalarH =0$ . It then follows that $\curl \sigma_{ab} = \curl\nhalf E_{ab} = H_{ab} =0$. The only remaining parts are therefore given by $\fscalarsigma$, $\fscalarE$ and $\fscalar{(\curlH)}$. 

If we now re-cast Eqs.\ (\ref{eeq})-(\r{evE}), in terms of the quantities that remain after symmetry restrictions, then we find
\bea
\dot{\Theta} &=& - \frac{1}{3} \Theta^2 -\frac32 ( \fscalarsigma)^2 \\
\fscalar{\dot{\sigma}} &=& - \frac{1}{3} \fscalarsigma \left( 2 \Theta +\frac{3}{2} \fscalarsigma \right) - \fscalarE \\
\fscalar{\nhalf \dot{E}} &=& - \fscalarE \left( \Theta -\frac{3}{2} \fscalarsigma \right) + \fscalar{(\curlH)} \, .
\eea
These equations can be further simplified by defining the expansion scalars parallel and perpendicular to $n^a$: $\mathcal{H}_\parallel := \frac{1}{3} \Theta+ \fscalarsigma$ and $\mathcal{H}_\perp := \frac{1}{3} \Theta -\frac{1}{2} \fscalarsigma$, respectively. The evolution equations for these two quantities are then given by
\bea
\l{ev1}
\dot{\mathcal{H}}_{\parallel} + (\mathcal{H}_{\parallel})^2 &=& - \fscalarE \\
\dot{\mathcal{H}}_{\perp} + (\mathcal{H}_{\perp})^2 &=& \frac{1}{2} \fscalarE \, ,
\l{ev2}
\eea
while the evolution equation for the source term, $\fscalarE$, is given by
\be
\l{ev3}
\fscalar{ \nhalf \dot{E}} +3 \mathcal{H}_{\perp} \fscalarE = \fscalar{(\curlH)} \, .
\ee
The scalar $\fscalar{(\curlH)}$ therefore acts as a source for $\fscalarE$, which is itself the source of the expansion along $\ss$. This completes our study of the restrictions imposed at reflection-symmetric surfaces. In the following sections we will study the evolution of the non-vanishing parts of the covariant scalars, and in particular the effect that $\fscalar{(\curlH)}$ has on the expansion of LRS curves with threefold, or higher, reflective symmetries.

\section{Taylor Series Expansion of Curl(H)}
\label{calcs}

We want to solve the evolution equations (\ref{ev1})-(\ref{ev3}), with the initial conditions specified in Section \ref{sec:models}. In terms of the variables that result from the $1+3$ decomposition, these initial conditions can be specified as
\be
\l{init1}
\Theta=0 =\sigma_{ab} = H_{ab} \qquad {\rm and} \qquad \dot{\Theta}=0 = \dot{H}_{ab}
\ee
together with the following conditions on $E_{ab}$:
\be
\l{init2}
(\div E)_a = 0 = (\curl E)_{ab} \, .
\ee
The first set of initial conditions in Eq.\ (\r{init1}) all follow immediately from the requirement of time-reversal symmetry in the initial data \c{Clifton_etal:2013}. The second set, involving time derivatives, follows from Eq.\ (\r{eeq}), and from the fact that the Cotton-York tensor on the initial hypersurface is given by ${}^*C_{ab} = - \dot{H}_{ab}$ \c{frames}. The conditions in Eq.\ (\r{init2}) then follow from the evolution equation (\r{evH}), and from the constraint equation (\r{c5}).

If one were to take $\fscalar{(\curlH)}=0$ along LRS curves, as was done in \c{Clifton_etal:2013}, then Eqs.\ (\r{ev1})-(\r{ev3}) form a closed system, and the initial data specified in Section \r{sec:models} is sufficient to determine $\mathcal{H}_{\parallel}$, $\mathcal{H}_{\perp}$ and $\fscalarE$ at all future times (up to singularities). If  $\fscalar{(\curlH)} \neq 0$, however, then this is no longer true. The value of $\fscalar{(\curlH)}$ must be determined at future times, in order to evolve the system of evolution equations (\r{ev1})-(\r{ev3}). This can be done by either numerically solving the full set of Einstein equations, or by performing a Taylor series expansion of $\fscalar{(\curlH)}$ at the initial hypersurface \c{KHB}. In this paper we opt for the latter of these two methods, and in the remainder of this section we determine the coefficients of such an expansion using the compact notation from Section \r{sec:formalism}.

In order to perform such an expansion, we first note that the time derivative of any spatial unit vector can be written as
\be
{\dot{e}_{1}}^a = {e_1}^{b} \dot{u}_b u^a - \Omega_3 {e_{2}}^a + \Omega_2 {e_{3}}^a \, ,
\ee
where $\Omega^{\alpha} := \frac{1}{2} \epsilon^{\alpha \beta \gamma} {e_{\beta}}^a \dot{e}_{\gamma a}$ is the angular velocity of the spatial unit vectors, with respect to a set that are Fermi propagated along $u^a$. If we choose our flow so that it is geodesic ($\dot{u}^a=0$), and take our unit vector $n^a$ to belong to a set of basis vectors that are Fermi-propagated ($\Omega^{\alpha}=0$), then we have $\dot{n}^a =0$ at all points along the evolution of all of our LRS curves. 

Letting $(\curl H_{ab})^{(n)}$ stand for the $n$th covariant derivative of $\curl H_{ab}$ along $u^a$, this means, in particular, that
\be
\fscalar{(\curlH)}^{(n)}   =   n^a n^a (\curl H_{ab})^{(n)} \, ,
\ee
for any number of time derivatives. We can therefore Taylor expand $\curl H_{ab}$, and simply contract the result with ${\bf n}$ twice, in order to find the series expansion of $\fscalar{(\curlH)}$.

\subsection{First Order}

The Taylor series expansion we wish to investigate is given by
\be
\l{taylor}
\curl H_{ab} = \sum_{n=0}^\infty \frac{(\curl H_{ab})^{(n)}\vert_{t=0}}{n!} \; t^n \, ,
\ee
where $(\curl H_{ab})^{(n)}$ should be understood to be evaluated on the initial hypersurface. The time parameter, $t$, is the proper time that would be measured by an observer following the integral curves of $u^a$.

The $n=0$ term in Eq.\ (\r{taylor}) is zero, as $H_{ab}$ must vanish at all points in the initial hypersurface. In fact, we can immediately see that every term that corresponds to an even value of $n$ must vanish from this equation. This follows directly from the requirement of time-reversal symmetry. It now remains to evaluate the odd values of $n$. At first order we get 
\begin{eqnarray}
\label{1ord}
({\rm curl}H_{ab})^{\centerdot} &=& -{\rm curl}(\Theta H_{ab})+3{\rm curl}(\sigma_{c\langle a}H_{b\rangle}{}^{c}) -{\rm curl}({\rm curl}(E_{ab}))\\&&+3H_{c\langle a}H_{b\rangle }{}^{c}-\frac13 \Theta \half {\rm curl}(H_{ab})-\sigma_{e}{}^{c}\epsilon_{cd(a}D^e H_{b)}{}^{d}\,, \nonumber
\end{eqnarray}
where we used the commutation relation between the $\curl$ operator and time derivatives (Eq.\ (A18) from \cite{Maartens:1997}):
\begin{eqnarray}
\label{A18}
({\rm curl}S_{ab})^{\centerdot}={\rm curl}\dot S_{ab}-\frac13 \Theta \half {\rm curl}S_{ab}-\sigma_{e}{}^{c}\epsilon_{cd(a}D^e S_{b)}{}^{d}+3H_{c\langle a} S_{b\rangle }{}^{c} \, ,
\end{eqnarray}
where $S_{ab}$ is any projected symmetric tracefree tensor, $S_{ab}=S_{\langle ab \rangle}$. The initial conditions in Eqs.\ (\r{init1}) and (\r{init2}) can then be seen to imply
\beq
({\rm curl}H_{ab})^{\centerdot} \vert_{t=0} =0 \, ,
\eeq
at all points on the initial hypersurface. This means that the leading-order term in Eq.\ (\r{taylor}) cannot occur at any lower value than $n=3$.

\subsection{Third Order}

In order to find the third-order term in the Taylor series expansion, from Eq.\ (\ref{taylor}), we must differentiate each term in Eq.(\ref{1ord}) twice.  For the first term we get
\beq
-({\rm curl}(\Theta H_{ab}))^{\centerdot\centerdot}=-(\Theta \half {\rm curl} H_{ab}+\epsilon_{cd(a}H_{b)}{}^{d}D^c \Theta)^{\centerdot\centerdot}\, ,
\eeq
where we have again used the commutation rule from Eq.\ (\ref{A18}), as well as the following identity (Eq.\ (A16) from \cite{Maartens:1997}):
\beq
\label{A16}
{\rm curl}(f S_{ab})=f \half {\rm curl}( S_{ab})+\epsilon_{cd(a}S_{b)}{}^{d}D^c f \, ,
\eeq
where $f$ is any scalar function. Applying the Leibniz rule, and plugging in the initial conditions, we then get $-({\rm curl}(\Theta H_{ab}))^{\centerdot\centerdot} \vert_{t=0}=0$.

For the second term in Eq.(\ref{1ord}) we get
\begin{eqnarray}
3 ({\rm curl}(\sigma_{c\langle a}H_{b\rangle }{}^{c}))^{\centerdot\centerdot} &=& 3 \Big[ {\rm curl}(\sigma_{c\langle a}H_{b\rangle }{}^{c})^\centerdot -\frac13 \Theta \half {\rm curl}(\sigma_{c\langle a}H_{b\rangle }{}^{c})\\&&\qquad 
-\sigma_{e}{}^{c}\epsilon_{c}{}^{d}{}_{(a}D^e \sigma_{\vert g \vert \langle b)}H_{d\rangle }{}^{g}+3H^{c}{}_{\langle a}\sigma_{\vert d \vert \langle b\rangle }H_{c\rangle }{}^{d} \Big]^\centerdot \nonumber
\end{eqnarray}
where we twice applied the identity from Eq.\ (\ref{A18}), and where we have also used vertical bars around indices to exclude them from symmetrization operations. The initial conditions stated in Eq.\ (\ref{init1}) then imply that $({\rm curl}(\sigma_{c\langle a}H_{b\rangle }{}^{c}))^{\centerdot\centerdot} \vert_{t=0}=0$, as all terms contain quantities that vanish at $t=0$. The second time derivative of the fourth, fifth and sixth terms in Eq.(\ref{1ord}) can also be seen to vanish, using similar logic.

It is therefore only the third term on the right-hand side of Eq.\ (\r{1ord}) that can be non-zero, which means $({\rm curl}( H_{ab}))^{\centerdot\centerdot\centerdot}\vert_{t=0}=-({\rm curl}\,{\rm curl}( E_{ab}))^{\centerdot\centerdot}$. Commuting the $\curl$ operator with the time derivatives, using Eq.\ (\r{A18}) and the initial conditions from Eqs.\ (\ref{init1}) and (\r{init2}), then gives
\be
\l{ccEdd}
-({\rm curl}\,{\rm curl}( E_{ab}))^{\centerdot\centerdot} \vert_{t=0}=-{\rm curl}\,{\rm curl}\ddot E_{ab}  - {\rm curl}(\epsilon_{cd(a}D^e (E_{b)}{}^{c} E_{e}{}^{d})) \, .
\ee
This can be simplified further by using the following identity (Eq.\ (A4) from \cite{Maartens:1997}):
\beq
\l{A4}
{\rm curl} (S^2)_{ab}=\epsilon_{cd(a}D^e (S_{b)}{}^{c} S_{e}{}^{d}) \, ,
\eeq
where $(S^2)_{ab}:=S_a^{\p{a}c} S_{bc}$. We then have, after using Eqs.\ (\r{ccEdd})-(\r{A4}) and the evolution equations (\ref{evs})-(\ref{evE}), that $-({\rm curl}\,{\rm curl}( E_{ab}))^{\centerdot\centerdot} \vert_{t=0}=4 \half {\rm curl}\, {\rm curl}(E^2)_{ab}$. This is the only term that is non-zero, after twice differentiating the right-hand side of Eq.\ (\ref{1ord}).

The final result, for the third derivative of $\curl H_{ab}$ on the initial hypersurface, can therefore be written as
\be
\l{cHddd}
({\rm curl}H_{ab})^{\centerdot\centerdot\centerdot} \vert_{t=0}=4 \half {\rm curl}\, {\rm curl}\left( E^2 \right)_{ab}\, .
\ee
This is the first non-zero coefficient in the Taylor series from Eq.\ (\ref{taylor}). Explicitly evaluating $({\rm curl}H_{ab})^{\centerdot\centerdot\centerdot}$, in terms of functions that appear in the metric, is likely to result in a very long expression, for all but the most trivial geometries. It is therefore remarkable that the same quantity can be written down in such a simple way in Eq.\ (\ref{cHddd}). This is a result of the compact notation that we have used, which hides an enourmous amount of underlying complexity.

\subsection{Fifth Order}

The third-order coefficient, found in the previous section, is expected to be the leading-order term in the Taylor series expansion from Eq.\ (\r{taylor}), as it is the first one that does not vanish. In order to determine the regime in which the Taylor series provides a good approximation to the full geometry, however, it is necessary to find the next-to-leading order term (when the leading-order and next-to-leading-order terms become the same size, then we expect the series expansion to break down). We therefore need to calculate the fifth-order term in the series expansion from Eq.\ (\r{taylor}). 

The fifth-order term will be given by performing four differentiations on each term in Eq.(\ref{1ord}). We will do this by following the same procedure used in the previous section, where simplifications were obtained using the identities in Eqs.\ (\r{A18}), (\r{A16}) and (\r{A4}), and the initial conditions in Eqs.\ (\r{init1}) and (\r{init2}). Each of the six terms in Eq.\ (\r{1ord}) will be considered separately, and then summed to give out final result.

\subsubsection{First Term.} To find the fourth derivative of the first term in Eq.\ (\ref{1ord}), it is useful to first determine the second derivative without applying the initial conditions.  This is given by
\begin{eqnarray}
\hspace{-60pt}
&&-\left( {\rm curl}(\Theta H_{ab}) \right)^{\centerdot\centerdot} \nonumber \\
&=& -\ddot \Theta {\rm curl} H_{ab} -2\dot \Theta {\rm curl} \dot H_{ab}+\frac13 \Theta \dot \Theta {\rm curl} H_{ab}+\dot \Theta \sigma_{e}{}^{c} \epsilon_{cd(a}D^e H_{b)}{}^{d} \nonumber\\
&&- 3\dot \Theta  H_{c\langle a}H_{b\rangle }{}^{c}-\Theta{\rm curl}\ddot H_{ab}+\frac13 \Theta^2 {\rm curl}\dot H_{ab}+\Theta \sigma_{e}{}^{c} \epsilon_{cd(a}D^e \dot H_{b)}{}^{d} \nonumber\\
&&- 3\Theta  H_{c\langle a } \dot H_{b\rangle }{}^{c}+\frac23 \Theta \dot \Theta  {\rm curl} H_{ab} +\frac13 \Theta^2  {\rm curl} \dot H_{ab}-\frac19 \Theta^3  {\rm curl} H_{ab} \nonumber\\
&&- \frac13 \Theta^2 \sigma_{e}{}^{c} \epsilon_{cd(a}D^e H_{b)}{}^{d} +\Theta^2  H_{c\langle a}H_{b\rangle }{}^{c}+\dot \Theta \sigma_{e}{}^{c} \epsilon_{cd(a}D^e H_{b)}{}^{d} \nonumber\\
&&+ \Theta \sigma_{e}{}^{c} \epsilon_{cd(a} (D^e  \dot H_{b)}{}^{d} - \frac13 \Theta D^e  H_{b)}{}^{d} - \sigma^{ed}D_{|d|} H_{b)}{}^{d}+2 H^{ec} \epsilon^{a}{}_{\vert e \vert \langle b)} H_{d\rangle }{}^{e})\nonumber\\
&&+ \Theta \dot \sigma_{e}{}^{c} \epsilon_{cd(a}D^e H_{b)}{}^{d}  - 3 \dot \Theta  H_{c\langle a}H_{b\rangle }{}^{c} -6 \Theta  H_{c\langle a } \dot H_{b\rangle }{}^{c} \, ,
\end{eqnarray}
where we first expanded ${\rm curl}(\Theta H_{ab})$ using Eq.\ (\ref{A16}), and then applied the commutation relation from Eq.\ (\ref{A18}) twice. It can be seen from this equation that, if we differentiate twice more, then each term will contain at least one factor of a quantity that vanishes on the initial hypersurface. We therefore have
\beq
-{\rm curl}(\Theta H_{ab})^{\centerdot\centerdot\centerdot\centerdot} \vert_{t=0}=0 \, ,
\eeq
so that the first term of Eq.\ (\ref{1ord}) does not contribute to the fifth derivative of $\curl H_{ab}$.

\subsubsection{Second Term.} Applying Eq.\ (\r{A18}), it can be seen that the second derivative of the second term in Eq.\ (\ref{1ord}) is given by
\begin{eqnarray}
3[{\rm curl}(\sigma_{c\langle a}H_{b\rangle }{}^{c})]^{\centerdot\centerdot}&=& [3 {\rm curl}((\sigma_{c\langle a}H_{b\rangle }{}^{c})^\centerdot)- \Theta {\rm curl}(\sigma_{c\langle a}H_{b\rangle }{}^{c}) \nonumber\\
&&\quad - 3\sigma_{e}{}^{g}\epsilon_{gd(a} D^e (\sigma_{|c|\langle b}H^{d\rangle c})+9H_{d\langle a}\sigma_{|c|\langle b \rangle}H^{d\rangle c}]^\centerdot \, .
\end{eqnarray}
This allows us to see, by inspection, that it is only the first term in this equation that will survive two further differentiations. After commuting these derivatives with the $\curl$ operator, and disregarding terms that vanish on the initial hypersurface, we then arrive at $3[{\rm curl}(\sigma_{c\langle a}H_{b\rangle }{}^{c})]^{\centerdot\centerdot\centerdot\centerdot} \vert_{t=0}= 3 \half {\rm curl} ((\sigma_{c\langle a}H_{b\rangle }{}^{c})^{\centerdot\centerdot\centerdot\centerdot})$. The only part of $(\sigma_{c\langle a}H_{b\rangle }{}^{c})^{\centerdot\centerdot\centerdot\centerdot}$ that does not vanish in the initial data is given by $\dot \sigma_{c\langle a} ( H_{b\rangle }{}^{c})^{\centerdot\centerdot\centerdot}$. Using the evolution equations for $\sigma_{ab}$ and $H_{ab}$, and the result of the previous section, in Eq.\ (\r{cHddd}), then gives
\be
3[{\rm curl}(\sigma_{c\langle a}H_{b\rangle }{}^{c})]^{\centerdot\centerdot \centerdot\centerdot}\vert_{t=0} = -48 {\rm curl}[E_{c\langle a}{\rm curl}(E^2)_{b\rangle }{}^{c}] \, .
\ee
This is the simplest expression we can find for this term.

\subsubsection{Third Term.} This is the most lengthy term to evaluate. We find, by repeated application of the commutation relation (\r{A18}), that
\begin{eqnarray}
\hspace{-60pt}
&&-[{\rm curl}( {\rm curl}  E_{ab})]^{\centerdot\centerdot} \nonumber \\
&=&-{\rm curl}(({\rm curl} E_{ab})^{\centerdot\centerdot}  )+\frac13 \Theta \half {\rm curl}(({\rm curl} E_{ab})^{\centerdot}  )+\sigma_{e}{}^{c} \epsilon_{cd(a}D^e( ({\rm curl} E_{b)}{}^{d})^\centerdot)  \nonumber\\
&&-3 H_{c\langle a}({\rm curl} E_{b\rangle }{}^{c})^\centerdot +\frac13 (\Theta {\rm curl}({\rm curl} E_{ab}))^\centerdot \nonumber\\
&&+ (\sigma_{e}{}^{c}\epsilon_{cd(a}D^e ({\rm curl} E_{b)}{}^{d}))^\centerdot - 3(H_{c\langle a}{\rm curl} E_{b\rangle }{}^{c})^\centerdot \, .
\end{eqnarray}
The complete evaluation of the third term requires two further derivatives, and we can see that the only terms that survive this operation are given by
\begin{eqnarray}
\hspace{-20pt}
-[{\rm curl}( {\rm curl}  E_{ab})]^{\centerdot\centerdot\centerdot\centerdot} \vert_{t=0}
&=&  - {\rm curl}(( {\rm curl}  E_{ab})^{\centerdot\centerdot\centerdot\centerdot}) + 6 \dot \sigma_{e}{}^{c}\epsilon_{cd(a} D^e (({\rm curl} E_{b)}{}^{d})^{\centerdot\centerdot}) \, , \l{ccEdddd}
\end{eqnarray}
where we have made use of the initial conditions. In order to evaluate this term we note that, after some manipulation, we can write
\begin{eqnarray}
( {\rm curl}  E_{ab})^{\centerdot\centerdot\centerdot\centerdot} \vert_{t=0}
&=&  - 8  {\rm curl} (E^2 E_{ab}) + 30 {\rm curl}(E^3)_{ab}+ 4 {\rm curl}\, {\rm curl}\,{\rm curl} (E^2)_{ab} \nonumber\\
&&- 9  E_{e}{}^{c} \epsilon_{cd(a} D^e   (E^2)_{b)}{}^{d} + 2  (E^2)_{e}{}^{c} \epsilon_{cd(a} D^e   E_{b)}{}^{d} \nonumber\\
&&- 9  E_{e}{}^{c} \epsilon_{cd(a} D^e   (E^2)_{b)}{}^{d} + 12  E^{c}{}_{\langle a} {\rm curl}(E^2)_{b\rangle c} \,, \l{cEdddd}
\end{eqnarray}
where we have again made repeated use of Eq.\ (\ref{A18}), as well as the following identity (Eq.\ (A12) from \cite{Maartens:1997}):
\be
\left( D_a S_{bc} \right)^{\centerdot} = D_a \dot{S}_{bc} - \frac{1}{3} \Theta D_a S_{bc} - \sigma_a^{\p{a} d} D_d S_{bc} +2 H_{a}^{\p{a} d} \epsilon_{de (b}^{} S_{c)}^{\p{c)} e} \, .
\ee
We have also made the definition $(S^3)_{ab} := S_{a c} S^{cd} S_{db}$. Substituing Eq.\ (\ref{cEdddd}) into Eq.\ (\ref{ccEdddd}), and performing some further manipulations, then gives
\begin{eqnarray}
\hspace{-70pt}
&&-[{\rm curl}( {\rm curl}  E_{ab})]^{\centerdot\centerdot\centerdot\centerdot} \nonumber \\
&=& 8 {\rm curl}\,{\rm curl}(E^2  E_{ab})- 30{\rm curl}\,{\rm curl}(E^3)_{ab} +18 \epsilon_{cd(a} {\rm curl}( D^e (E_{\vert e \vert}{}^{c} (E^2)_{b)}{}^{d}) \nonumber\\
&&- 2 \epsilon_{cd(a} {\rm curl} ((E^2)_{\vert e \vert}{}^{c}D^e  E_{b)}{}^{d}) -12  {\rm curl}(E^{c}{}_{\langle a} {\rm curl} (E^2)_{b\rangle c})\nonumber\\
&& - 4 {\rm curl}\,{\rm curl}\,{\rm curl}\,{\rm curl}(E^2)_{ab} +24\epsilon_{cd(a} D^e (  E_{e}{}^{c} {\rm curl}  ( E^2)_{b)}{}^{d}) \, .
\end{eqnarray}
This is the lengthiest of the six terms being evaluated.

\subsubsection{Fourth and Fifth Terms.} The fourth and fifth terms, from Eq.\ (\r{1ord}), can both be seen to give
\beq
3(H_{c\langle a}H_{b\rangle }{}^{c})^{\centerdot\centerdot\centerdot\centerdot} \vert_{t=0} =0 \, ,
\ee
and
\beq
-\frac13 (\Theta {\rm curl}(H_{ab}))^{\centerdot\centerdot\centerdot\centerdot} \vert_{t=0}=0 \, .
\eeq
This is because $H_{ab}$ and $\Theta$, together with both their first and second derivatives, must vanish on the initial hypersurface. Two further time derivatives would therefore be required before either of these terms become non-zero at $t=0$.

\subsubsection{Sixth Term.} The final term evaluates to
\begin{eqnarray}
-[\sigma_{e}{}^{c}\epsilon_{cd(a}D^e H_{b)}{}^{d}]^{\centerdot\centerdot\centerdot\centerdot} \vert_{t=0}
&=& 16  E_{e}{}^{c}\epsilon_{cd(a}D^e [{\rm curl}(E^2)_{b)}{}^{d} ] \, ,
\end{eqnarray}
where we have used the commutation rules from Eqs.\ (\r{A18}) and (\r{A16}), as well as the evolution equations (\r{evs}), (\r{evE}) and (\r{evH}).

Putting all of these results together leads us to our final expression for the fifth order term in our Taylor series expansion of $\curl H_{ab}$:
\begin{eqnarray}
\hspace{-60pt}
&&(\curl (H)_{ab} )^{\centerdot\centerdot\centerdot\centerdot\centerdot} \nonumber \\
&=& 8 \curl \curl \left( E^2 E_{ab} \right) -30 \curl \curl \left( (E^3)_{ab} \right) -60 \curl \left( E^c_{\phantom{c} \langle a} \curl (E^2)_{b \rangle c} \right) \nonumber\\
&&+18  \curl \left( \epsilon_{cd(a} D^e \left( (E^2)_{b)}^{\phantom{b)} d}  E_{e}^{\phantom{e} c} \right) \right) -2  \curl \left( (E^2)_{e}^{\phantom{e} c} \epsilon_{cd(a}  D^e E_{b)}^{\phantom{b)} d}  \right) \nonumber \\
&&+40 \epsilon_{cd (a} D^e \left(  \curl (E^2)_{b)}^{\phantom{b)} d} E_{e}^{\phantom{e} c} \right) -4 \curl \curl \curl \curl (E^2)_{ab} \, . \l{cHddddd}
\end{eqnarray}
This equation is considerably less compact than the result for the third-order derivative, presented in Eq.\ (\r{cHddd}), but is still remarkably short, given the complexity it conceals.

In what follows, we will use the expressions calculated in Eqs.\ (\r{cHddd}) and (\ref{cHddddd}) to evaluate the first two non-vanishing terms in the Taylor series from Eq.\ (\ref{taylor}). This will allow us to not only determine the leading-order effect that $H_{ab}$ has on the expansion of LRS curves, but also to estimate when the series expansion approach breaks down. We will perform this analysis for all three classes of LRS curves, and in all six of the lattices that were described in Section \ref{sec:models}.

\section{The Effect of Curl(H) on Cosmology}
\l{sec:results}

The effect that $H_{ab}$ has on the expansion of space can be estimated by considering its consequences on the scale factors, $a_{\parallel}$ and $a_{\perp}$, defined implicitly by
\be
\mathcal{H}_{\parallel} = \frac{\dot{a}_{\parallel}}{a_{\parallel}} \qquad {\rm and} \qquad
\mathcal{H}_{\perp} = \frac{\dot{a}_{\perp}}{a_{\perp}} \, .
\label{scales}
\ee
The function $a_{\parallel}$ is then the scale factor along our LRS curves, and $a_{\perp}$ is the scale factor in all perpendicular directions. Both are a function of proper time, along the integral curves of ${\bf u}$, as well as position in space.

Following the method used in \cite{KHB}, and extending it to higher order, we can use Eqs.\ (\ref{ev1})-(\ref{ev3}) to write the correction to the scale factors, from including the $\curl H_{ab}$ term in Eq. (\ref{ev3}), as
\begin{eqnarray}
\Delta a_{\parallel} &=& 
- \frac{ a_{\parallel} \half \fscalar{(\curlH)}^{\centerdot\centerdot\centerdot}\vert_{t=0}}{ 6!} \half t^6 \\ \nonumber
&&+\frac{ a_{\parallel}}{8!} \left[ 25  \half \fscalar{(\curlH)}^{\centerdot\centerdot\centerdot} \half \fscalar{E}-  \fscalar{(\curlH)}^{\centerdot\centerdot\centerdot \centerdot\centerdot} \right] \Big|_{t=0} \half t^8 +O(t^{10})  
\l{da1}
\end{eqnarray}
and
\begin{eqnarray}
\Delta a_{\perp} &=&
\frac{ a_{\perp} \half \fscalar{(\curlH)}^{\centerdot\centerdot\centerdot}\vert_{t=0}}{2 \times 6!} \half t^6 \\ \nonumber
&&-\frac{ a_{\perp}}{2 \times 8!} \left[  \fscalar{(\curlH)}^{\centerdot\centerdot\centerdot} \half \fscalar{E}-  \fscalar{(\curlH)}^{\centerdot\centerdot\centerdot \centerdot\centerdot} \right]\Big|_{t=0} \half t^8 +O(t^{10}) \, ,
\l{da2}
\end{eqnarray}
where all quantities on the right-hand side of these equations should be taken to be evaluated at $t=0$. These expressions result from simultaneously Taylor expanding $a_{\parallel}$, $a_{\perp}$ and $\fscalar{(\curl H)}$ about the initial hypersurface, and substituting the results into Eqs.\ (\ref{ev1})-(\ref{ev3}). The value of $\fscalar{E}$ is then given by noting that on a time-symmetric hypersurface, and in vacuum, the electric part of the Weyl tensor is equal to the Ricci tensor of the 3-space, $E_{ab} = {}^{(3)}R_{ab}$. The quantities involving derivatives of $\curl H_{ab}$ are given, in terms of $E_{ab}$ and its derivatives, by Eqs.\ (\ref{cHddd}) and (\r{cHddddd}). We therefore have all the information required to evaluate all of the terms present in Eqs.\ (\r{da1}) and (\r{da2}).

The LRS curves that we will consider in this section are those displayed in Fig.\ref{cubefig}. The red curve in this figure extends halfway along a cell edge, from the centre of an edge to a vertex. The green curve connects the horizon of the mass at the centre with the centre of a cell face, and the blue curve connects the horizon and a cell vertex. These curves are depicted in Fig.\ \ref{cubefig} for the special case of a cubic cell, which is the basic constituent of the 8-cell. However, similar curves also exist for the case of tetrahedral, octahedral and dodecahedral cells. We will not present specific diagrams of these cells, but instead rely on the reader to visualize the corresponding curves in each of these cases.

The proper length of each of our LRS curves, in a hypersurface of constant proper time, is then given by the integral
\be
L(t) = \int_{\chi_1}^{\chi_2} a_{\parallel} (t, \chi) d \chi \, ,
\l{length}
\ee
where $\chi$ is the coordinate from Eq.\ (\ref{sol1}), and where we have rotated the configuration so that the LRS curve under consideration is at constant $\theta$ and $\phi$. In what follows, we will present the intial values of the functions $\fscalar{E}$, as well as the derivatives $\fscalar{(\curl H)}^{\centerdot\centerdot\centerdot}$ and $\fscalar{(\curl H)}^{\centerdot\centerdot\centerdot\centerdot\centerdot}$, at every point along the three LRS curves depicted in Fig.\ \ref{cubefig}. We will then calculate the proper length of each of these curves, as a function of proper time along the integral curves of ${\bf u}$, until the Taylor expansion given in Eq.\ (\ref{taylor}) breaks down. In calculating these results, we have substituted the expansion from Eq.\ (\ref{taylor}) directly into the evolution equations (\r{ev1})-(\r{ev3}). 

The indicator that we use to determine when the Taylor series approximation breaks down is when the first two non-vanishing terms on the right-hand side of Eq.\ (\r{taylor}) become equal. This happens at different times along each of the curves, depending on the spatial position that one considers. We choose the two endpoints of each of the curves, as depicted in Fig.\ \ref{cubefig}, in order to follow the magnitude of the ratio of these two terms. This is a convenient choice as the endpoints of the curve are uniquely picked out by the geometry of the problem. We also expect these points to give a fair reflection of the validity of the Taylor expansion at intermediate points, as they are in the two most extreme environments available along each curve (in the case of the green and blue curves, one end touches a black hole while the other is equidistant from black holes). If the value of $\fscalar{(\curl H)}^{\centerdot\centerdot\centerdot}$ or $\fscalar{(\curl H)}^{\centerdot\centerdot\centerdot\centerdot\centerdot}$ vanishes at any one of these points, then we simply take the ratio of the relevant terms a very small distance away (1\% of the distance along the curve, where all quantities are non-zero).

\subsection{Curves Along Cell Edges}

The first set of curves we wish to consider are those that lie along the edges of our cells, as depicted by the red line in Fig.\ \ref{cubefig}. In this study we will take the curve in question to begin at the cell vertex, and extend halfway along the length of the edge. We will then calculate the length of this curve, using the method described above, until the Taylor series approximation has broken down at both of its ends (i.e. at both a point near the cell vertex, and at the centre of a cell edge).

\begin{figure}[h!]
\begin{centering}
\includegraphics[width=3.5in]{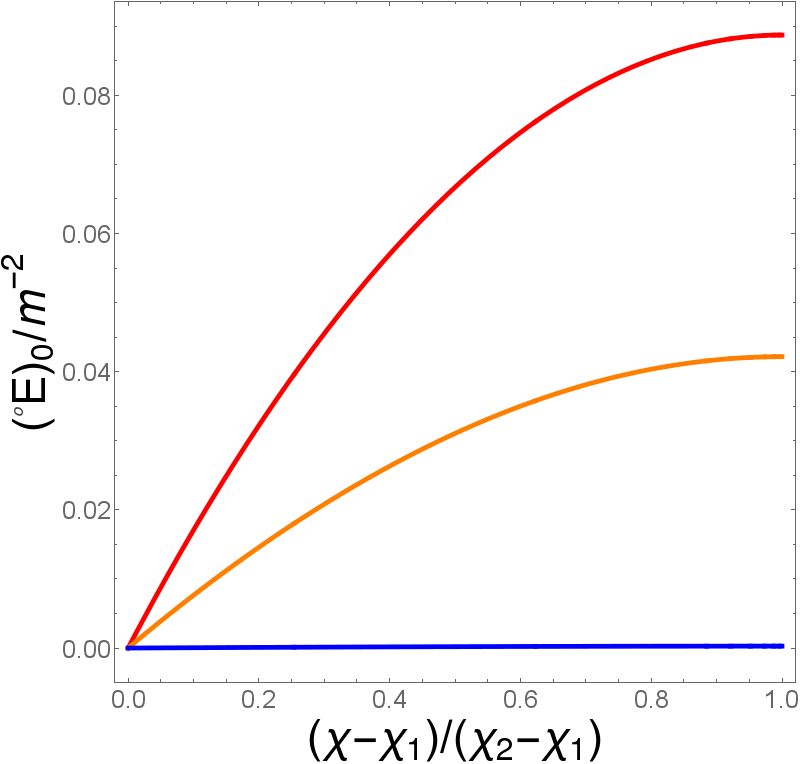}
\par\end{centering}
\caption{The value of $\fscalar{E}$ at $t=0$ along a cell edge for the 5-cell (red), the 8-cell (orange), and the 120-cell (blue). The proper mass within each cell is denoted $m$.}
\centering{}\label{efig1}
\begin{centering}
\includegraphics[width=3.5in]{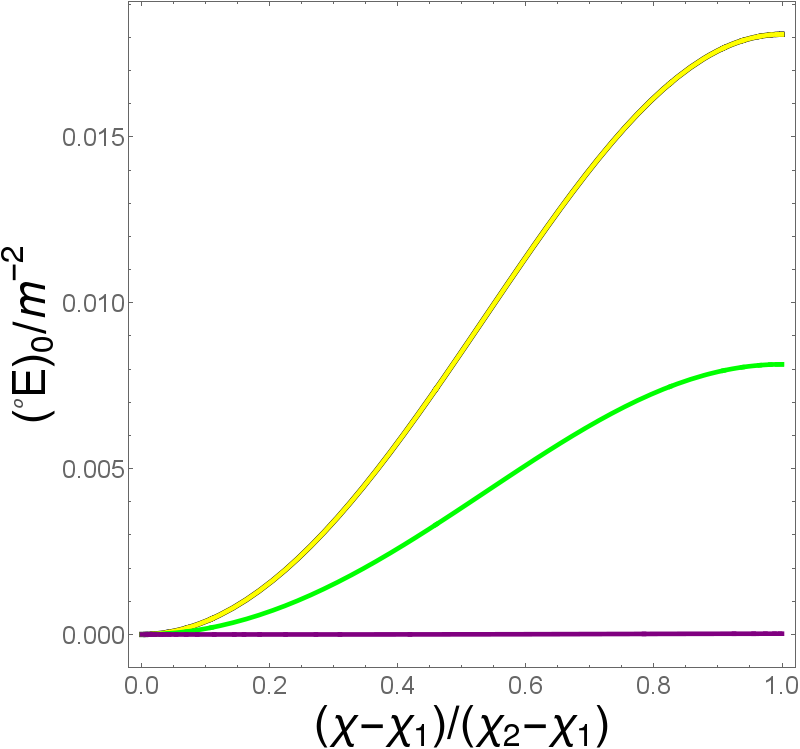}
\par\end{centering}
\caption{The value of $\fscalar{E}$ at $t=0$ along a cell edge for the 16-cell (yellow), the 24-cell (green), and the 600-cell (purple). The proper mass within each cell is $m$.}
\centering{}\label{efig2}
\end{figure}

First let us present the relevant information about the electric and magnetic parts of the Weyl tensor along these curves, and on the initial hypersurface. The form of the electric part of the Weyl tensor, contracted twice with the space-like unit vector tangent to the edge, is displayed in Figs. \ref{efig1} and \ref{efig2}. In these plots we have chosen the cell vertex to be located at $\chi= \chi_1$, and the centre of the cell edge to be located at $\chi=\chi_2$. This information is presented in two different plots, as there are two different functional forms for $\fscalar{E}$; those that have a non-zero derivative at $\chi=\chi_1$, and those that have a vanishing derivative at $\chi=\chi_1$. These two cases correspond to lattices in which the cell edges are non-contiguous and contiguous, respectively. The former of these two cases contains the 5-cell, the 8-cell and the 120-cell lattices, while the latter contains the 16-cell, the 24-cell and the 600-cell lattices.

\begin{figure}[h!]
\begin{centering}
\includegraphics[width=3.4in]{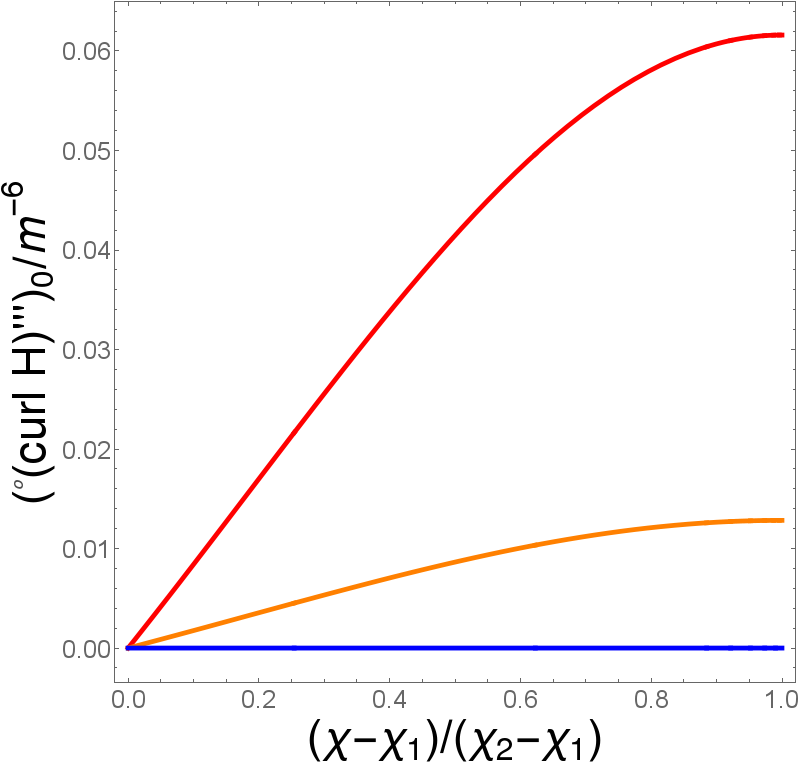}
\par\end{centering}
\caption{The value of $\fscalar{(\curl H)}^{\centerdot \centerdot \centerdot}$ at $t=0$ along a cell edge for the 5-cell (red), the 8-cell (orange), and the 120-cell (blue). The proper mass within each cell is $m$. The blue curve in this plot is positive valued.}
\centering{}\label{efig3}
\begin{centering}
\includegraphics[width=3.5in]{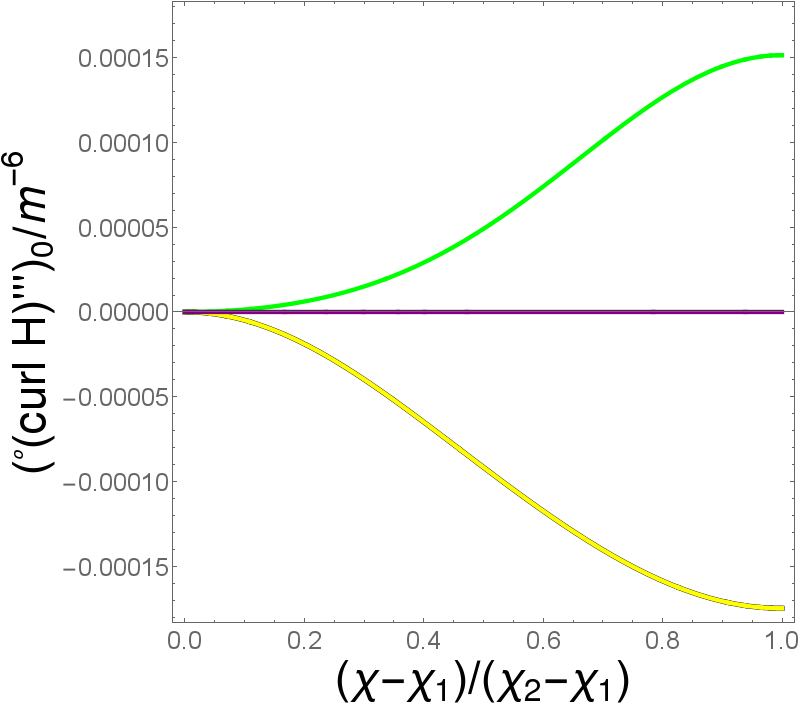}
\par\end{centering}
\caption{The value of $\fscalar{(\curl H)}^{\centerdot \centerdot \centerdot}$ at $t=0$ along a cell edge for the 16-cell (yellow), the 24-cell (green), and the 600-cell (purple). The proper mass within each cell is $m$. The purple curve is negative valued.}
\centering{}\label{efig4}
\end{figure}

\begin{figure}[h!]
\begin{centering}
\includegraphics[width=3.4in]{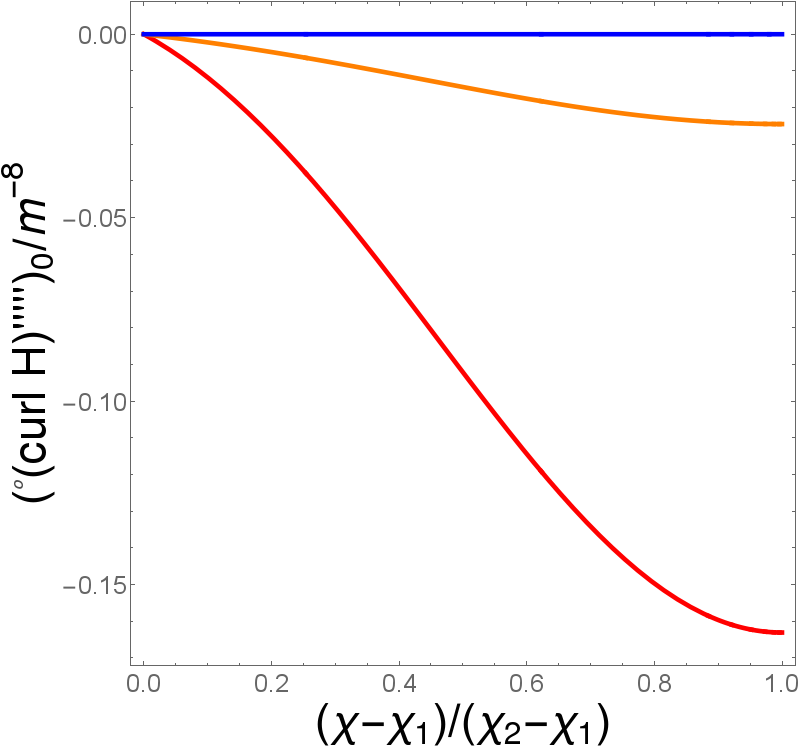}
\par\end{centering}
\caption{The value of $\fscalar{(\curl H)}^{\centerdot  \centerdot \centerdot \centerdot \centerdot}$ at $t=0$ along a cell edge for the 5-cell (red), the 8-cell (orange), and the 120-cell (blue). The proper mass within each cell is $m$. The blue line is negative valued.}
\centering{}\label{efig5}
\begin{centering}
\includegraphics[width=3.4in]{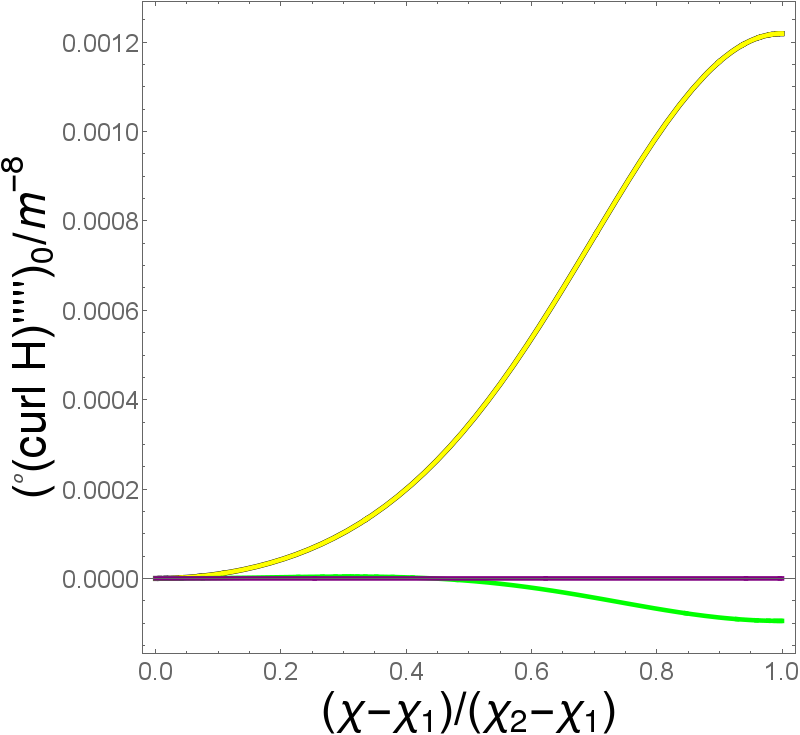}
\par\end{centering}
\caption{The value of $\fscalar{(\curl H)}^{\centerdot \centerdot \centerdot \centerdot \centerdot}$ at $t=0$ along a cell edge for the 16-cell (yellow), the 24-cell (green), and the 600-cell (purple). The proper mass within each cell is $m$. The purple line is negative valued.}
\centering{}\label{efig6}
\end{figure}

As well as the electric part of the Weyl tensor, we also need to know about the magnetic part of the Weyl tensor. This tensor vanishes on the initial hypersurface, but its derivatives do not. In Figs. \ref{efig3}-\ref{efig6} we present information about the value of the these derivatives, evaluated on the initial hypersurface at each point along the curves under consideration. Figs. \ref{efig3} and \ref{efig4} were produced by twice contracting the result from Eq.\ (\r{cHddd}) with the space-like tangent vector ${\bf n}$. Figs. \ref{efig5} and \ref{efig6} were produced using Eq.\ (\r{cHddddd}). We again display these results for the lattices with contiguous and non-contiguous edges separately, as they have different functional forms. Interestingly, although the curves have similar shapes to those of $\fscalar{E}$, they do not all have the same sign (the curve for the 16-cell can be seen to be negative in Fig.\ \r{efig4}, and positive in Fig.\ \r{efig6}). One may also note that the magnitude of $\fscalar{(\curl H)}^{\centerdot \centerdot \centerdot}$ and $\fscalar{(\curl H)}^{\centerdot \centerdot \centerdot \centerdot \centerdot}$ are both much smaller than $\fscalar{E}$, in the cases where the lattices have contiguous edges (and in the chosen units).

Next, we use the information presented in Figs. \ref{efig1}-\ref{efig6} to evolve Eqs.\ (\r{ev1})-(\r{ev3}). The lengths of the curves being considered in this section can then be calculated by finding the scale factors defined in Eq.\ (\r{scales}), and integrating them as prescribed in Eq.\ (\ref{length}). If the scale factor becomes zero at any point along the curve, we simply remove this point from the integral, as described in \cite{Clifton_etal:2013}. The result of all of this is shown in Fig.\ \r{efig7}. The proper length of the curve is evolved in time, in this plot, and displayed using units of the total proper mass in the cosmology (i.e. the proper mass of each cell, multipled by the number of cells in the lattice). This choice of units allows us to display all six lattices together with just a single FLRW solution that has the same total mass as each of the lattices. The particular FLRW solution we choose to compare with our models is a dust-filled cosmology with positive spatial curvature.

\begin{figure}[t!]
\begin{centering}
\includegraphics[width=4in]{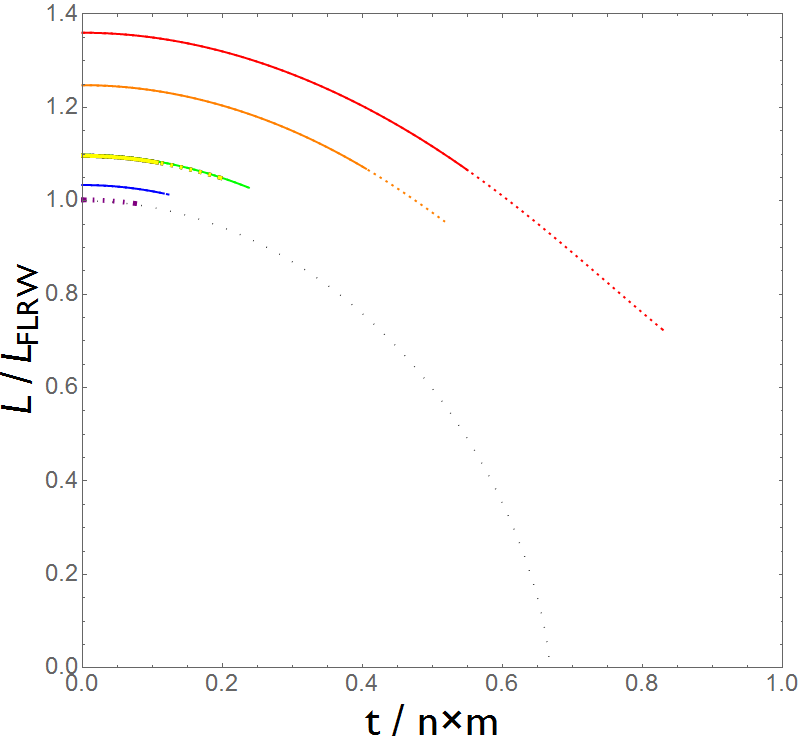}
\par\end{centering}
\caption{The length of the cell edge, normalized by the maximum value of a corresponding curve in an FLRW universe that is filled with dust, and has positive spatial curvature. The six different lattice models are displayed using the same colours as in Figs. \r{efig1}-\r{efig6}, and the FLRW solution with the same total mass is displayed as a black dotted line.}
\centering{}\label{efig7}
\end{figure}

\begin{figure}[h!]
\begin{centering}
\includegraphics[width=4.5in]{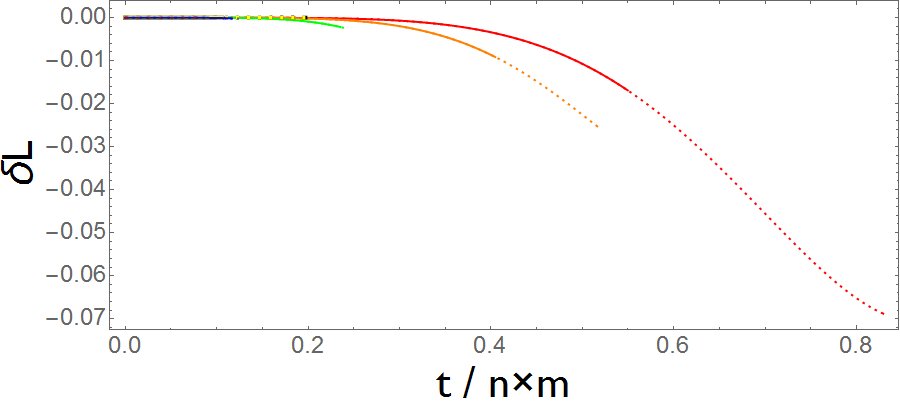}
\par\end{centering}
\caption{The difference between including $\fscalar{(\curl H)}$ in Eq.\ (\ref{ev3}) and neglecting it. This is displayed by showing the effect it has on the length of the cell edge, again normalized by the maximum value of a corresponding curve in an FLRW universe. The six different lattice models are displayed using the same colours as in Figs. \r{efig1}-\r{efig6}.}
\centering{}\label{efig8}
\end{figure}

\begin{figure}[t!]
\begin{centering}
\includegraphics[width=4in]{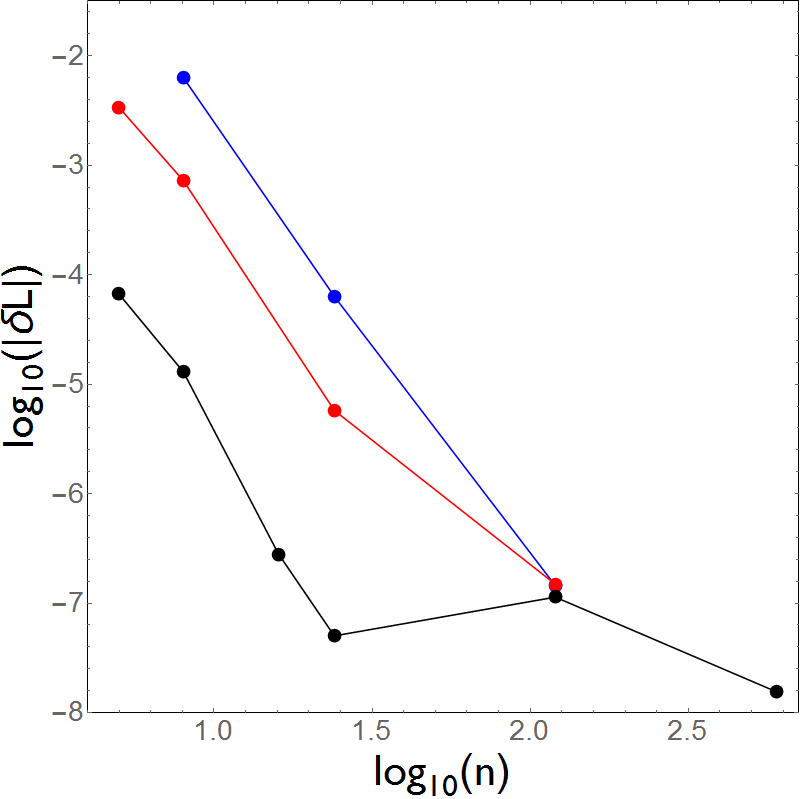}
\par\end{centering}
\caption{The difference between including $\fscalar{(\curl H)}$ in Eq.\ (\ref{ev3}) and neglecting it, as inferred from the length of a cell edge, and as a function of the number of cells in the lattice, $n$. This is shown at $t=m$ (black points), $t=2 m$ (red points), and $t=3 m$ (blue points), where $m$ is the proper mass contained within any one cell in the lattice. Points are excluded when the curves shown in Figs. \ref{efig7} and \ref{efig8} become dashed, rather than solid.}
\centering{}\label{efig9}
\end{figure}

Also indicated in Fig.\ \ref{efig7} is the region where the Taylor series expansion of $\fscalar{(\curl H)}$ breaks down. We indicate this on the plot by the solid line turning into a dashed line. We start the dashing when one end of the curve fails our convergence criterion ({\it i.e.} when the second non-zero term in Eq.\ (\ref{taylor}) becomes as large as the first). We end the dashing when the other end of the curve also fails this condition. The dashed region therefore corresponds to a domain where some of the points on the curve obey the convergence criterion, but not all. In this region the Taylor series approximation is breaking down. It can be seen that the length of the solid and dashed region, for each line on the plot, is different from the others. Roughly speaking, the Taylor series expansion seems to break down more quickly as the number of cells in the lattice increases.

As well as considering the evolutions of the overall length of the cell edge, which are quite similar to those found in \cite{Clifton_etal:2013}, it is also interesting to consider the magnitude of the effect of including $\fscalar{(\curl H)}$ in Eq.\ (\ref{ev3}). This effect is displayed graphically in Fig.\ \ref{efig9}, where we plot the difference between the curves shown in Fig.\ \ref{efig7} and the curves that would have existed if we had neglected $\fscalar{(\curl H)}$ (as was done in \cite{Clifton_etal:2013}). Once again, we used solid and dashed lines to indicate where the Taylor series expansion starts to break down. The effect of including $\fscalar{(\curl H)}$ is most dramatic in the smallest lattices. In the 5-cell and 8-cell the difference can reach values of $\vert \delta L \vert \simeq 0.01$. In the larger lattices this difference is much less, although the Taylor series expansion breaks down much sooner. One may also note, however, that at the same cosmological time (measured in units of $n \times m$) the value of $\delta L$ is larger in the bigger lattices.

As the curves in Fig.\ \ref{efig8} are difficult to distinguish, we have read off particular values for each of the lattices, and displayed these separately in a log-log-plot in Fig.\ \ref{efig8}. The three curves in this plot show the value of $\delta L$ at $t=m$ (black points), at $t=2 m$ (red points), and at $t=3 m$ (blue points), where $m$ is the proper mass in any given cell. We have excluded points that correspond to locations on the lines in Figs. \ref{efig7} and \ref{efig8} that become dashed, so as to exclude regions where the series expansion of $\fscalar{(\curl H)}$ breaks down. Although the black curve dips in the middle, the general trend seems to show that the effect of $\fscalar{(\curl H)}$ becomes smaller as the number of cells in the lattice is increased (at least, when comparing the lengths of the curves at these values of $t$). The effect of $\fscalar{(\curl H)}$ does, however, increase with time, reaching levels of $\sim 1\%$, in the smaller lattices at $t=3 m$.

\subsection{Curves Through Face Centres}

\begin{figure}[t!]
\begin{centering}
\includegraphics[width=4in]{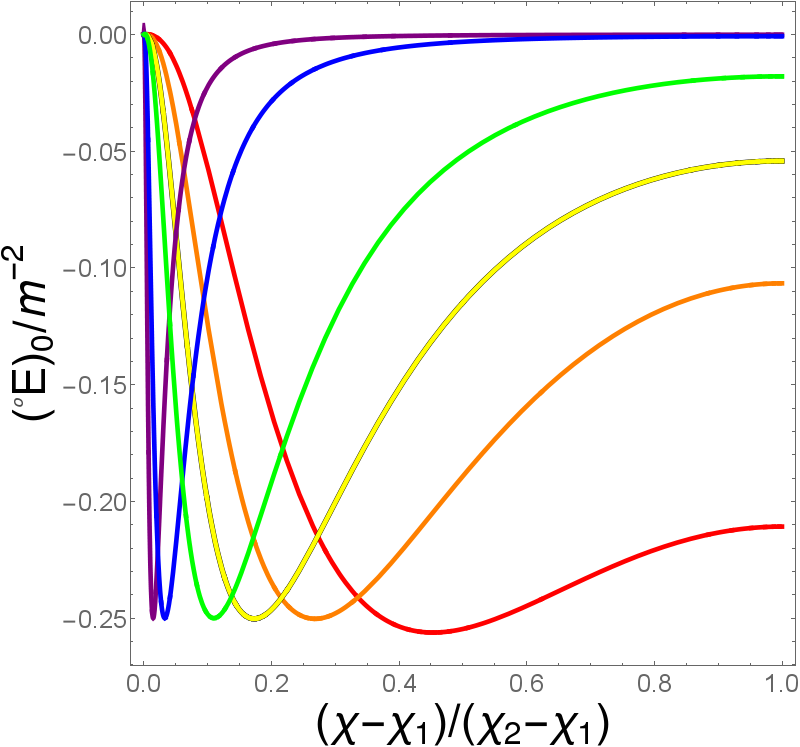}
\par\end{centering}
\caption{The value of $\fscalar{E}$ at $t=0$ along a curve that goes from the cell centre to the centre of a cell face, for the 5-cell (red), the 8-cell (orange), the 16-cell (yellow), the 24-cell (green), the 120-cell (blue), and the 600-cell (purple).}
\centering{}\label{cfig1}
\end{figure}

\begin{figure}[h!]
\begin{centering}
\includegraphics[width=3.5in]{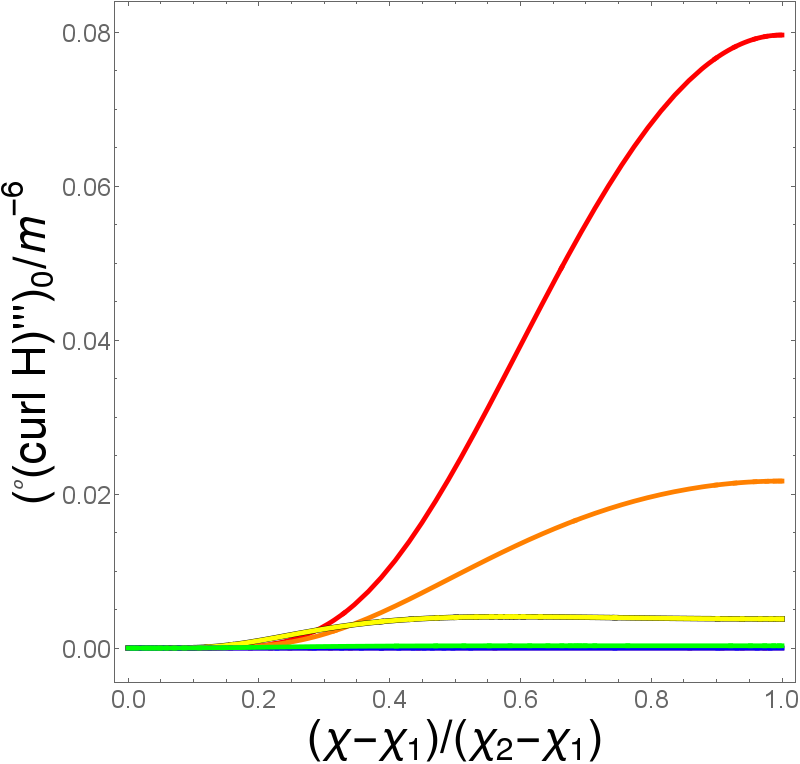}
\par\end{centering}
\caption{The value of $\fscalar{(\curl H)}^{\centerdot \centerdot \centerdot}$ at $t=0$ along a curve that goes from the cell centre to the centre of a cell face, for the 5-cell (red), the 8-cell (orange), the 16-cell (yellow), the 24-cell (green), the 120-cell (blue), and the 600-cell (purple).}
\centering{}\label{cfig2}
\begin{centering}
\includegraphics[width=3.6in]{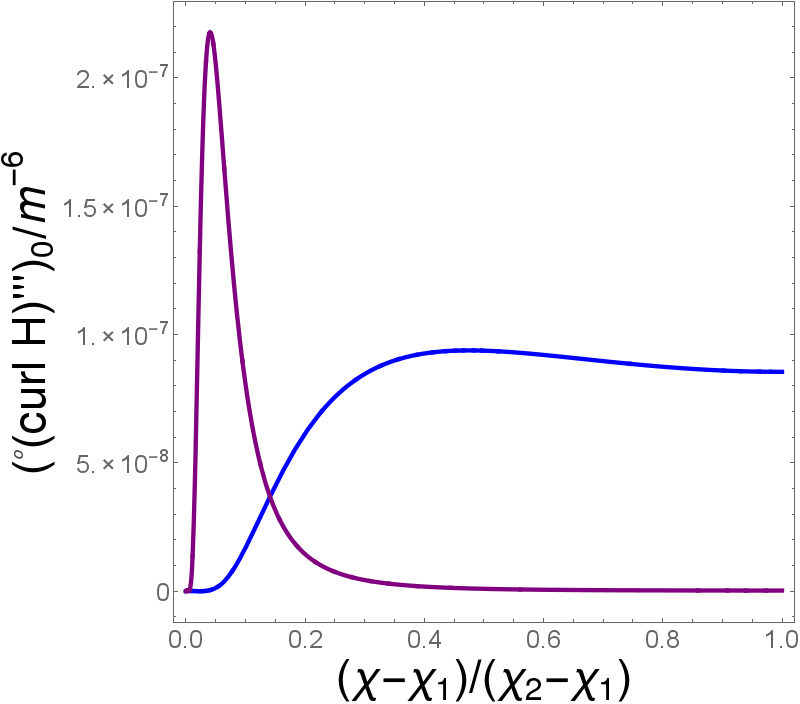}
\par\end{centering}
\caption{As in Fig.\ \r{cfig2}, but only showing the 120-cell (blue) and the 600-cell (purple).  Again, $m$ is the proper mass of the black hole at the centre of the cell.}
\centering{}\label{cfig3}
\end{figure}

\begin{figure}[h!]
\begin{centering}
\includegraphics[width=3.5in]{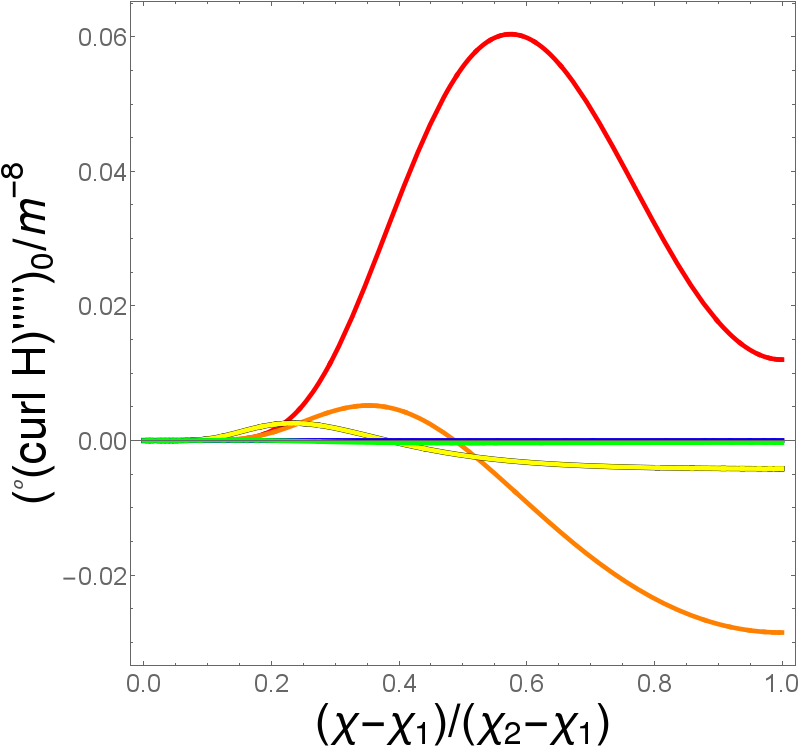}
\par\end{centering}
\caption{The value of $\fscalar{(\curl H)}^{\centerdot \centerdot \centerdot \centerdot \centerdot}$ at $t=0$ along a curve that goes from the cell centre to the centre of a cell face, for the 5-cell (red), the 8-cell (orange), the 16-cell (yellow), the 24-cell (green), the 120-cell (blue), and the 600-cell (purple).}
\centering{}\label{cfig4}
\begin{centering}
\includegraphics[width=3.6in]{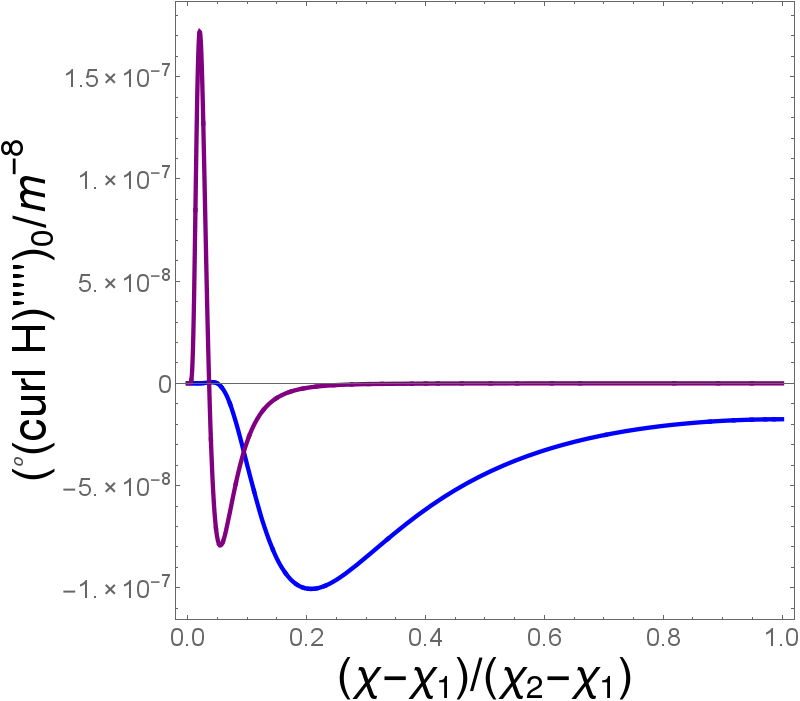}
\par\end{centering}
\caption{As in Fig.\ \r{cfig4}, but only showing the 120-cell (blue) and the 600-cell (purple). Once more, $m$ is the proper mass of the black hole at the centre of the cell.}
\centering{}\label{cfig5}
\end{figure}

Let us now set aside the curves at the edges of cells, and instead consider those that extend from the horizon of a black hole to the centre of a cell face. These are the type of curves depicted by the green line in Fig.\ \ref{cubefig}, for the particular case of a cubic cell. Such curves exhibit local rotational symmetry, in exactly the same way as a cell edge. Once again, we will evaluate $\fscalar{E}$ and the derivatives of $\fscalar{(\curl H)}$ along these curves. The former of these quantities is shown in Fig.\  \ref{cfig1}, while the latter are shown in Figs. \ref{cfig2}-\ref{cfig5}.

\begin{figure}[t!]
\begin{centering}
\includegraphics[width=4in]{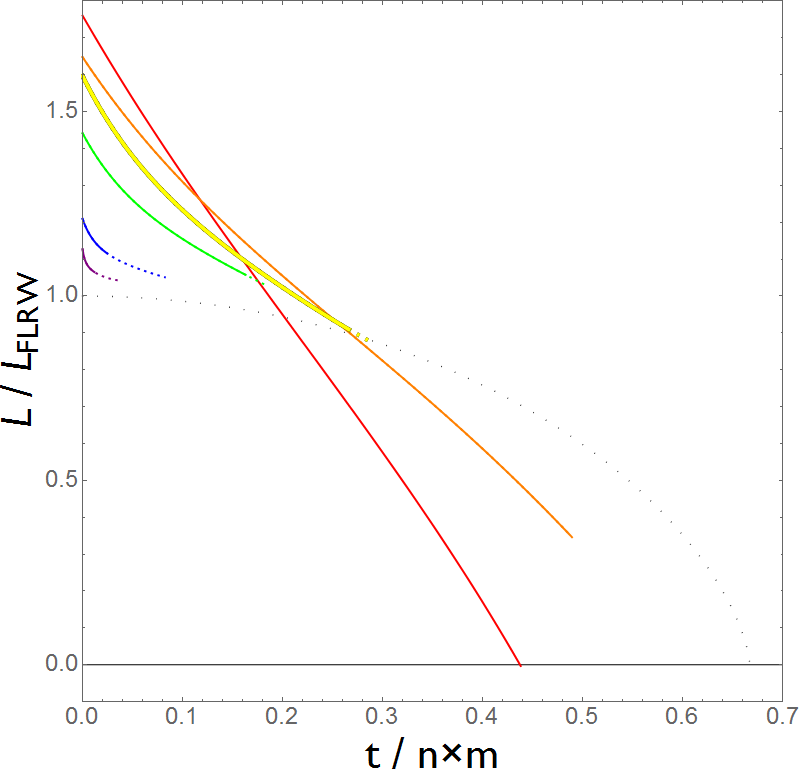}
\par\end{centering}
\caption{The length of the curve that goes from the cell centre to the centre of a cell face, normalized by the maximum value of a corresponding curve in an FLRW universe that is filled with dust, and has positive spatial curvature. The six different lattice models displayed as before, black dotted line is the corresponding FLRW solution.}
\centering{}\label{cfig6}
\end{figure}

\begin{figure}[t!]
\begin{centering}
\includegraphics[width=4.5in]{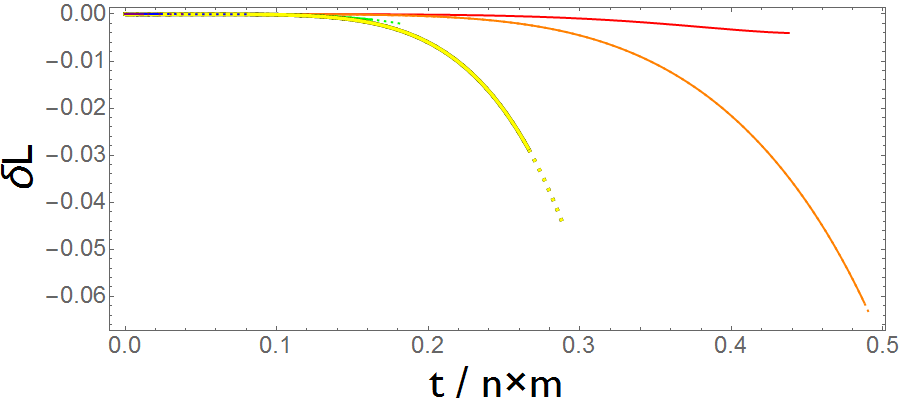}
\par\end{centering}
\caption{The difference between including $\fscalar{(\curl H)}$ in Eq.\ (\ref{ev3}) and neglecting it. This is displayed by showing the effect it has on the length of the curve that connects the horizon with the centre of a cell face. Colours denote lattices, as in Fig.\ \ref{cfig1}.}
\centering{}\label{cfig7}
\end{figure}

\begin{figure}[t!]
\begin{centering}
\includegraphics[width=4in]{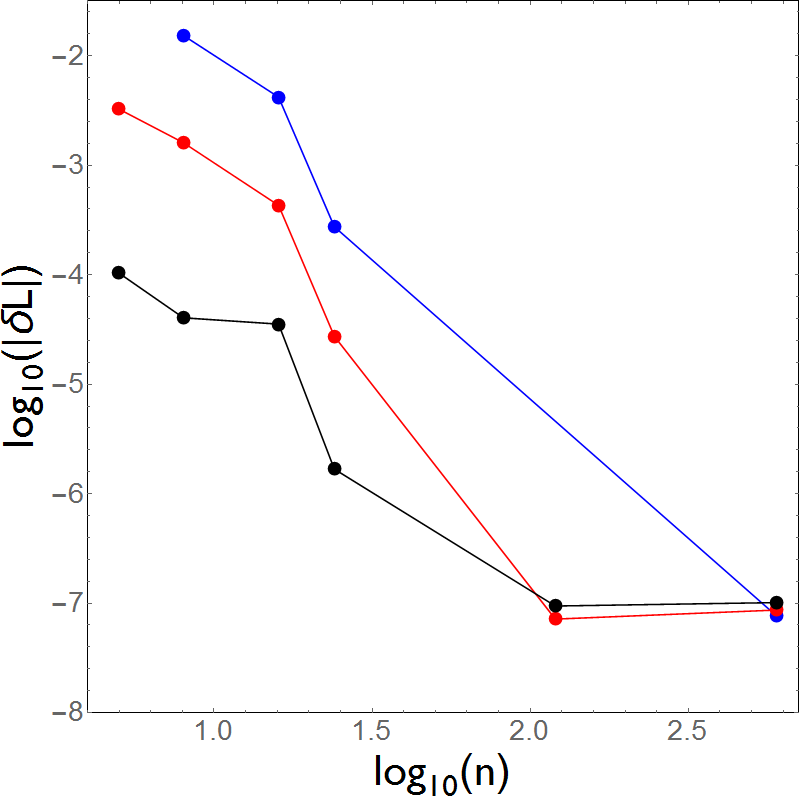}
\par\end{centering}
\caption{The difference between including $\fscalar{(\curl H)}$ in Eq.\ (\ref{ev3}) and neglecting it, for the curve that connects the horizon to the centre of a cell face, presented as a function of the number of cells in the lattice, $n$. This is shown at $t=m$ (black points), $t=2 m$ (red points), and $t=3 m$ (blue points), where $m$ is the proper mass contained within any one cell in the lattice. Points are excluded when the curves shown in Figs. \ref{cfig6} and \ref{cfig7} become dashed, rather than solid.}
\centering{}\label{cfig8}
\end{figure}

In these plots we have taken $\chi_1$ to correspond to the centre of the cell (rather than the location of the horizon). The curves then extend through the horizon, which is initially located at the minimum of $\fscalar{E}$ \cite{Clifton_etal:2013}, and out to the edge of the cell, where they meet the centre of a cell face at $\chi=\chi_2$. If the proper lengths of these curves were to be calculated from $\chi_1$ to $\chi_2$, in the initial hypersurface, then we would find the result to be divergent. We therefore restrict our calculations of the proper length to the region of space exterior to the horizon (a finite quantity). To find the location of the horizon at all times after $t=0$ we simply propagate a null geodesic outwards from the initial location, along the LRS curve \cite{Clifton_etal:2013}.

The form of $\fscalar{(\curl H)}^{\centerdot \centerdot \centerdot}$ along each of these curves is shown in Fig.\ \ref{cfig2}. The larger lattices are difficult to make out on this scale, so we also show them separately in Fig.\  \ref{cfig3}. The form of these curves in more complicated than was the case for the cell edges. In the two smallest lattices it is apparent that there is a single maximum in the function, located at the centre of the cell face. For the larger lattices, the maximum is located somewhere between the horizon and the cell face centre, with the cell face centre itself becoming a local minimum. The behaviour of $\fscalar{(\curl H)}^{\centerdot \centerdot \centerdot \centerdot \centerdot}$ is even more complicated, and is shown in Fig.\ \ref{cfig4} for all lattices, and in Fig.\ \ref{cfig5} for the 120-cell and 600-cell alone. It can be seen that both maxima and minima of this function exist, along the curves under consideration.

Using these data, we can again integrate Eqs.\ (\ref{ev1})-(\ref{ev3}), and calculate the length of the curve (from the black hole horizon to the centre of the cell face) using Eq.\ (\ref{length}). The results of doing this are shown in Fig.\ \ref{cfig6}, for each of the six lattices. In this case the effect of the $\fscalar{(\curl H)}$ term in Eq.\ (\ref{ev3}) is again small, and the curves look very similar to their counterparts when this term is neglected \cite{Clifton_etal:2013}. Once more, we make the lines in
Fig.\ \ref{cfig6} dashed when one end fails our convergence criterion, and we stop plotting the line once both ends fail. Somewhat surprisingly, the curves in this case extend slightly further than was the case with the cell edges. This means the Taylor series approximation is slightly better along these curves, and for the 5-cell it appears to be good enough to evolve the curve all the way until its proper length vanishes (when the horizon makes its way to the centre of the cell face).

In order to visualize the effect of the $\fscalar{(\curl H)}$ term on the evolution of this curve, we plot the difference from its inclusion in Fig.\ \ref{cfig7}. In this case, the difference in the proper length of the curve can be as much as $\sim 6 \%$ of its initial length (in the case of the 8-cell), and $\sim 3\%$ for the 16-cell. This larger difference is, to some extent, a consequence of the Taylor series approximation lasting longer in this case, meaning that the cumulative effect of integrating Eq.\ (\ref{ev3}) with an extra term is larger. Again, however, the value of $\delta L$ for the larger lattices is too small to be seen in Fig.\ \ref{cfig7}. We therefore read off its value at $t=m$, at $t=2 m$, and at $t=3 m$ (where $m$ is the proper mass of each of the black holes). This information is displayed graphically in Fig.\ \ref{cfig8}.

As before, when presenting the values of $\delta L$ in the log-log-plot shown in Fig.\ \ref{cfig8}, we do not present points that would correspond to locations along the curve where the Taylor series approximtation has started to break down. There are, however, more points in the plot shown in Fig.\ \ref{cfig8} than there are in the one shown in Fig.\ \ref{efig9}. This is again due to the Taylor series approximation lasting longer in the present case than it did along the cell edges. On the other hand, the broad trends in the two plots are largely similar: The value of $\delta L$ decreases as the number of cells in the lattice is increased, and increases as the time from the initial hypersurface is increased. The largest  values of $\delta L$ in this plot are again at the level of $\sim 1\%$, for the smallest lattices at $t=3 m$.

\subsection{Curves Through Vertices}

\begin{figure}[h!]
\begin{centering}
\includegraphics[width=3.5in]{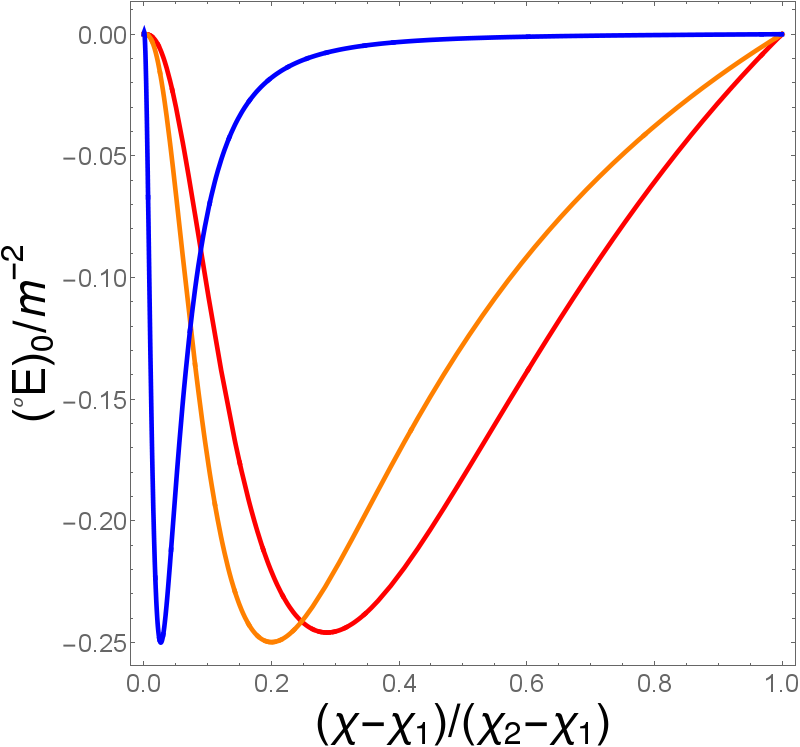}
\par\end{centering}
\caption{The value of $\fscalar{E}$ at $t=0$ along a diagonal curve for the 5-cell (red), the 8-cell (orange), and the 120-cell (blue).}
\centering{}\label{dfig1}
\begin{centering}
\includegraphics[width=3.5in]{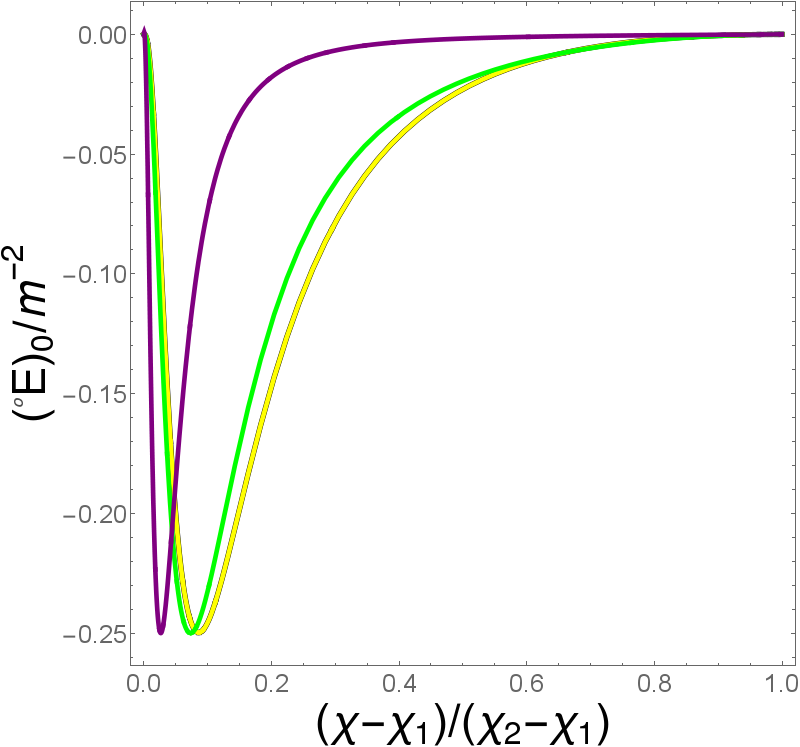}
\par\end{centering}
\caption{The value of $\fscalar{E}$ at $t=0$ along a diagonal curve for the 16-cell (yellow), the 24-cell (green), and the 600-cell (purple).}
\centering{}\label{dfig2}
\end{figure}

\begin{figure}[h!]
\begin{centering}
\includegraphics[width=3.6in]{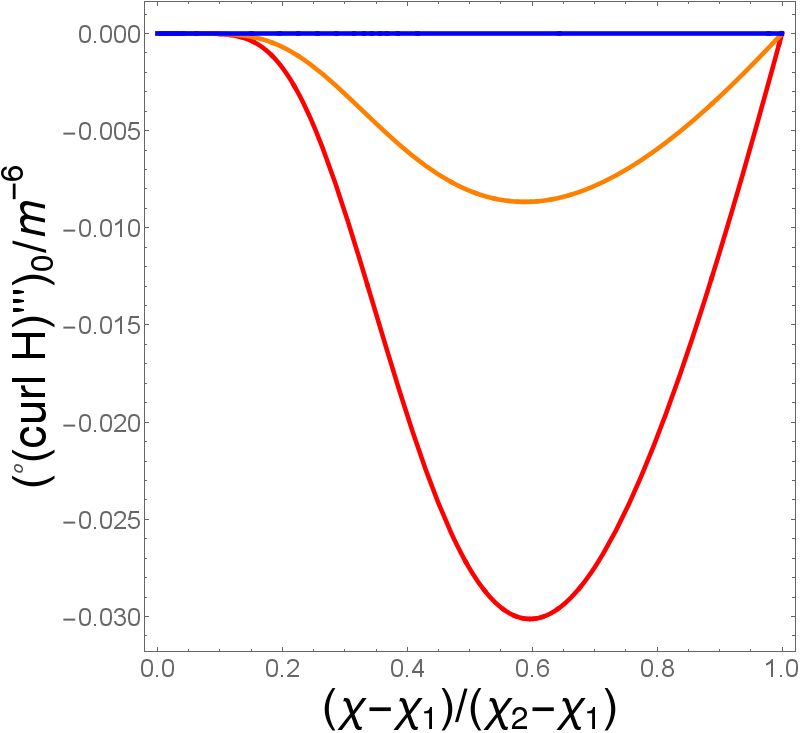}
\par\end{centering}
\caption{The value of $\fscalar{(\curl H)}^{\centerdot \centerdot \centerdot}$ at $t=0$ along a diagonal curve for the 5-cell (red), the 8-cell (orange), and the 120-cell (blue).}
\centering{}\label{dfig3}
\begin{centering}
\includegraphics[width=3.6in]{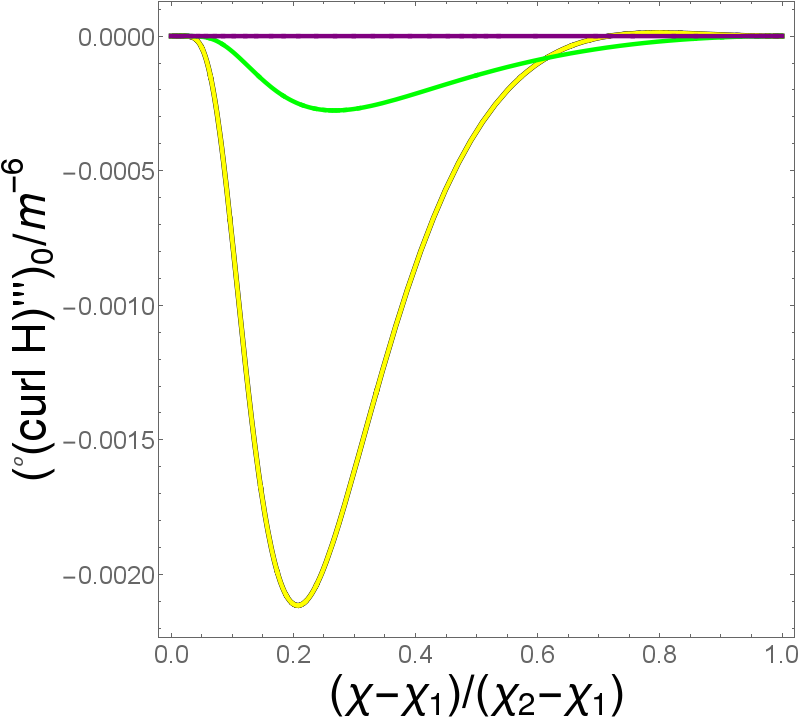}
\par\end{centering}
\caption{The value of $\fscalar{(\curl H)}^{\centerdot \centerdot \centerdot}$ at $t=0$ along a diagonal curve for the 16-cell (yellow), the 24-cell (green), and the 600-cell (purple).}
\centering{}\label{dfig4}
\end{figure}

\begin{figure}[h!]
\begin{centering}
\includegraphics[width=3.6in]{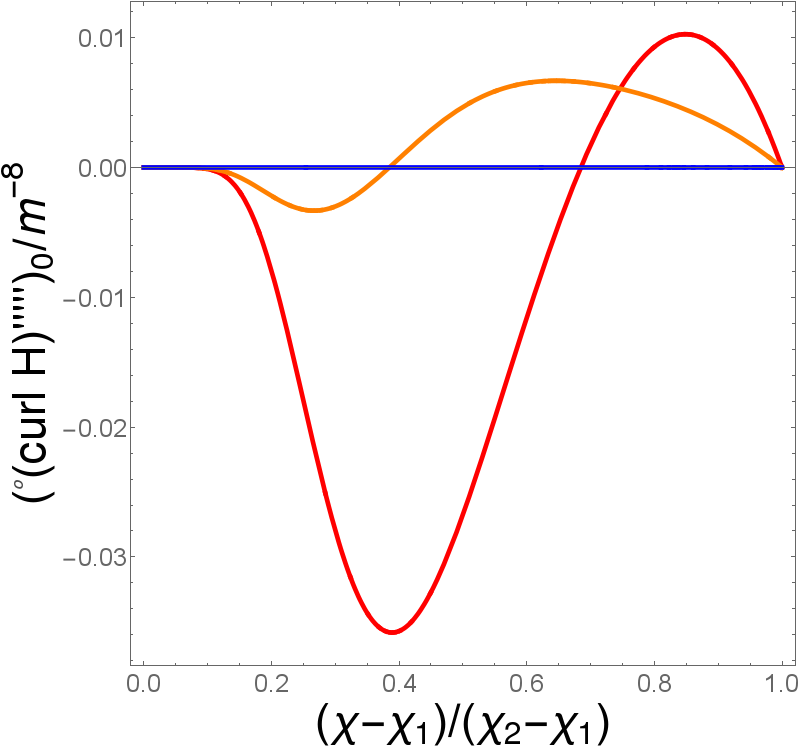}
\par\end{centering}
\caption{The value of $\fscalar{(\curl H)}^{\centerdot  \centerdot \centerdot \centerdot \centerdot}$ at $t=0$ along a diagonal curve for the 5-cell (red), the 8-cell (orange), and the 120-cell (blue).}
\centering{}\label{dfig5}
\begin{centering}
\includegraphics[width=3.6in]{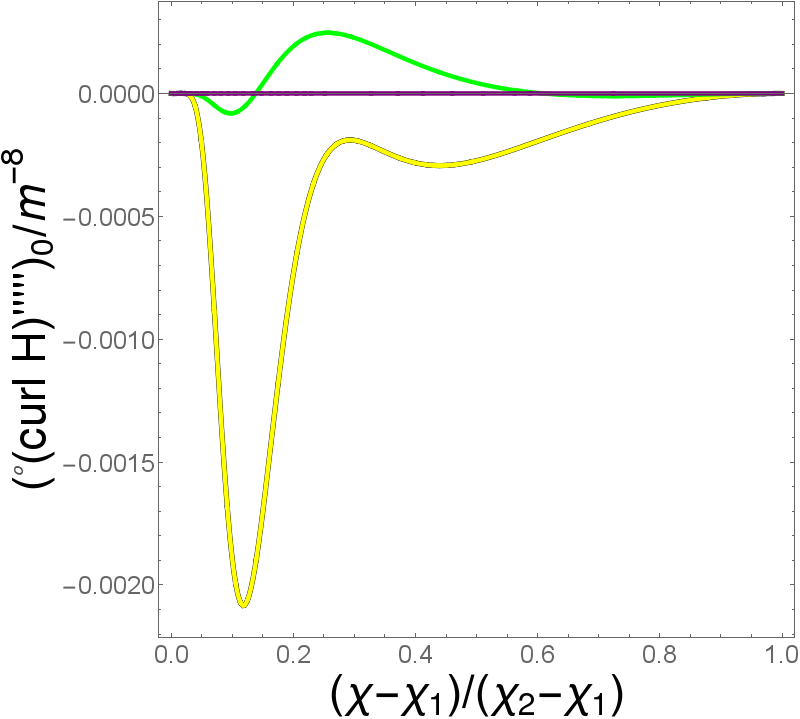}
\par\end{centering}
\caption{The value of $\fscalar{(\curl H)}^{\centerdot \centerdot \centerdot \centerdot \centerdot}$ at $t=0$ along a diagonal curve for the 16-cell (yellow), the 24-cell (green), and the 600-cell (purple).}
\centering{}\label{dfig6}
\end{figure}

\begin{figure}[t!]
\begin{centering}
\includegraphics[width=4in]{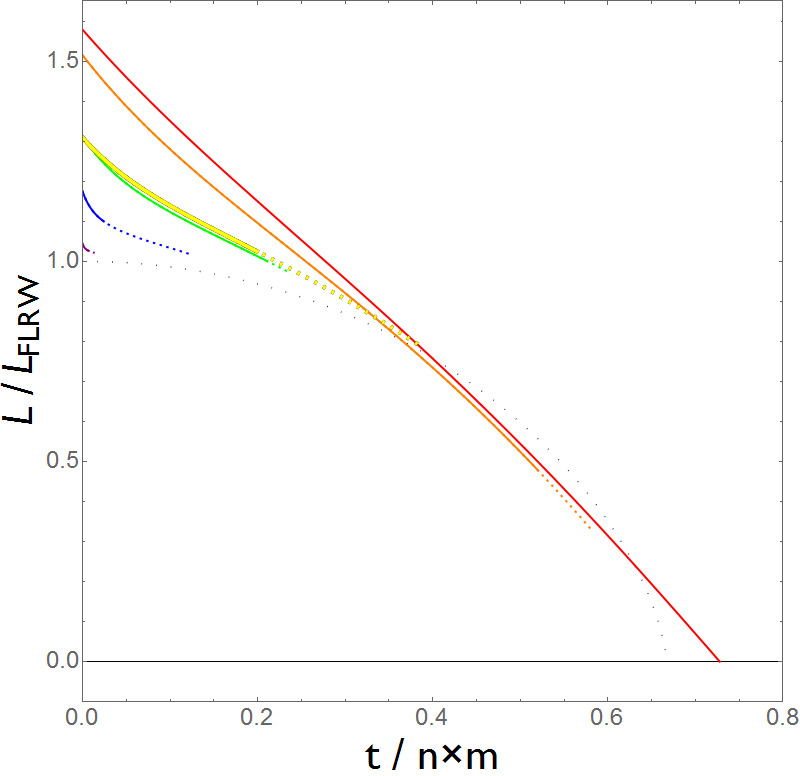}
\par\end{centering}
\caption{The length of a diagonal curve, normalized by the maximum value of a corresponding curve in an FLRW universe. The six different lattice models are displayed using the same colours as before, and the FLRW solution with the same total mass is displayed as a black dotted line.}
\centering{}\label{dfig7}
\end{figure}

\begin{figure}[t!]
\begin{centering}
\includegraphics[width=4.5in]{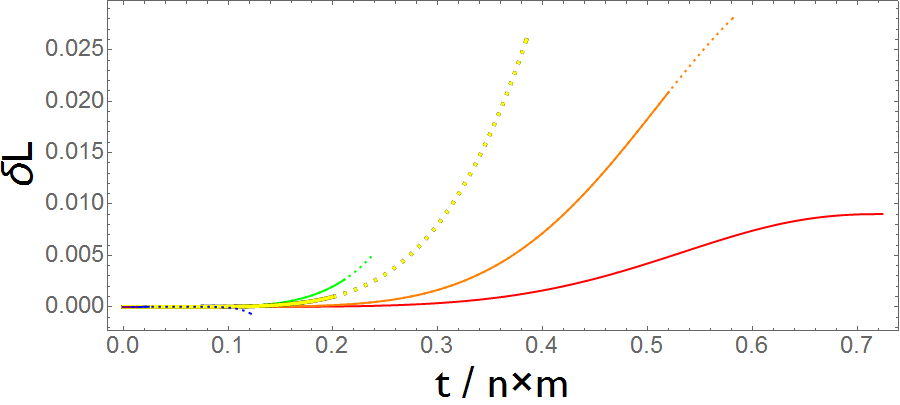}
\par\end{centering}
\caption{The difference between including $\fscalar{(\curl H)}$ in Eq.\ (\ref{ev3}) and neglecting it. This is displayed by showing the effect it has on the length of a diagonal curve, again normalized by the maximum value of a corresponding curve in an FLRW universe. The six different lattice models are displayed as before.}
\centering{}\label{dfig8}
\end{figure}

\begin{figure}[t!]
\begin{centering}
\includegraphics[width=4in]{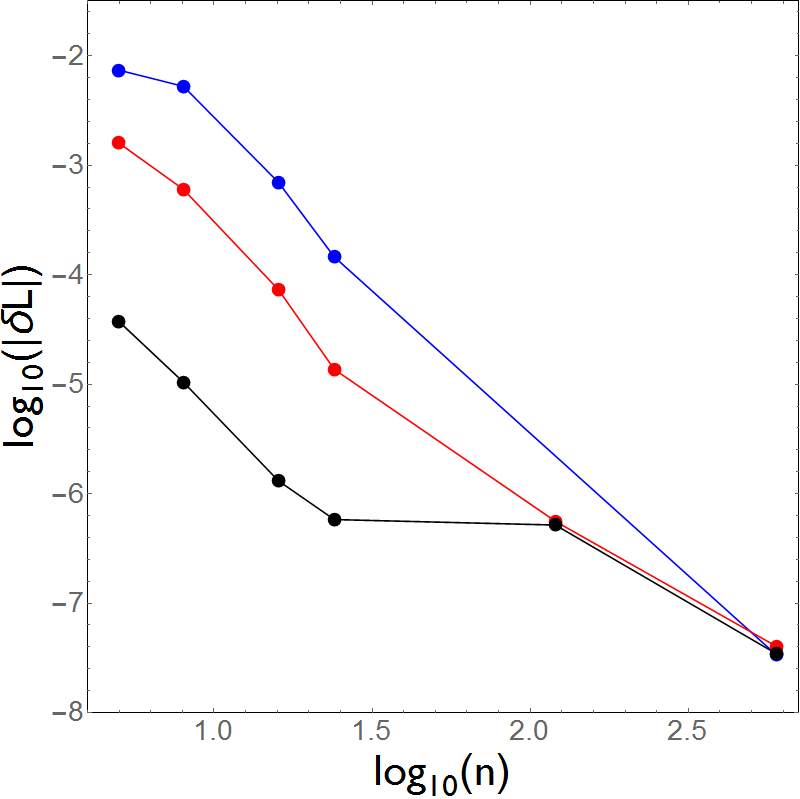}
\par\end{centering}
\caption{The difference between including $\fscalar{(\curl H)}$ in Eq.\ (\ref{ev3}) and neglecting it, as inferred from the length of a diagonal curve, and as a function of the number of cells in the lattice, $n$. This is shown at $t=m$ (black points), $t=2 m$ (red points), and $t=3 m$ (blue points), where $m$ is the proper mass contained within any one cell in the lattice. Points are excluded when the curves shown in Figs. \ref{dfig7} and \ref{dfig8} become dashed, rather than solid.}
\centering{}\label{dfig9}
\end{figure}

Finally, let us consider the remaining curves depicted in Fig.\ \ref{cubefig}: these are the blue curves that connect the black hole at the centre of the cell to one of the vertices. In the absence of any better terminology, we will call these curves `diagonals' (although this is probably only really a fitting description for the case of cubic cells). The location of the horizons will be determined in the same way as in the previous subsection. We find their initial position by looking for the minimum in the function $\fscalar{E}$, along our LRS curves, and then at subsequent times by propagating a null geodesic outwards from this location. In all of Figs. \ref{dfig1}-\ref{dfig6} we take the cell centre to correspond to the point $\chi=\chi_1$, and the cell vertex to correspond to the point $\chi=\chi_2$. We remind the reader that these curves are formally infinitely long, so we restrict our integrations to the regions of space exterior to the black hole horizons.

The values of $\fscalar{E}$ along the diagonals, for the six different lattices we are considering, are shown in Figs. \ref{dfig1} and \ref{dfig2}. The values of $\fscalar{(\curl H)}^{\centerdot \centerdot \centerdot}$ are shown in Figs. \ref{dfig3} and \ref{dfig4}, and the values of $\fscalar{(\curl H)}^{\centerdot \centerdot \centerdot \centerdot \centerdot}$ are shown in Figs. \ref{dfig5} and \ref{dfig6}. Just as with the cell edges, there are two different types of behaviour in this case, depending on whether or not the diagonal curve from one cell are contiguous with those of its neighbours, or not. It is still the case that the curves in the 5-cell, the 8-cell and the 120-cell are non-contiguous, while those of the 16-cell, the 24-cell and the 600-cell are contiguous. In the contiguous case one could image holding a mirror up to the right-hand side of the plots to determine the continuing form of the function in question, as the curve extends directly into the next cell. In the non-contiguous case the diagonal curve from one cell joins onto the end of the cell edge of its neighbours. One could therefore join the plots in Figs. (\ref{dfig1}), (\ref{dfig3}) and (\ref{dfig5}) with those shown in Figs. (\ref{efig1}), (\ref{efig3}) and (\ref{efig5}) to see the form of the functions in question if the diagonal curve were to be extended beyond the vertex.

While the functional form of $\fscalar{(\curl H)}^{\centerdot \centerdot \centerdot}$ along the diagonal curves is relatively simple, as can be seen from Figs. \ref{dfig3} and \ref{dfig4}, the form of $\fscalar{(\curl H)}^{\centerdot \centerdot \centerdot \centerdot \centerdot}$ is rather more complicated. Neverthelss, the evolutions equations (\ref{ev1})-(\ref{ev3}) can again be integrated, and used to calculate the proper length of the curves, as prescribed by Eq.\ (\ref{length}). The results of this are shown in Fig.\ \ref{dfig7}, and the difference plot showing the effect of including $\fscalar{(\curl H)}$ are shown in Fig.\ \ref{dfig8}. The magnitude of the effect is again most pronounced for the smaller lattices, as in these cases it takes longer for the Taylor series approximation to break down. Again, it is hard to see the effect of $\fscalar{(\curl H)}$ on the larger lattices, so we have read off values of $\delta L$ at $t=m$, at $t=2m$ and at $t=3 m$. These are shown in Fig.\ \ref{dfig9}, and can be seen to show the same broad trends as the other two sets of LRS curves, considered above. The magnitude of the effect of including $\fscalar{(\curl H)}$ is largest for the smallest lattices (when compared at the same time, as measured in units of $m$). It also grows as time goes on, and can reach the level of $\sim 1\%$ by the time $t= 3m$.

\section{Discussion}
\l{sec:discussion}

We have considered the effect of $H_{ab}$ on the evolution of LRS curves in lattice models of the universe. These models treat the matter in the Universe as a collection of point-like sources, and allow the formulation of cosmology as an initial value problem. They are therefore particularly well suited to the study of relativistic effects in cosmology, including the study of the evolution and effects of the magnetic part of the Weyl tensor.

We find that although the initial data of our models is silent (with $H_{ab}=0$), the evolution of the space is not silent. In particular, a $\curl H_{ab}$ term appears in the evolution equation for the electric part of the Weyl tensor, which in turn acts as the source for the evolution of LRS curves. This result was identified and studied numerically by Korzy\'{n}ski, Hinder and Bentivegna for the particular case of a universe that contains eight black holes \cite{KHB}. We extend their study by calculating the leading-order and next-to-leading order terms in a Taylor series approximation that can be used to incorporate the effects of $H_{ab}$ on LRS curves, and by applying it to all possible regular arrangements of black holes in a closed cosmological model. The inclusion of the next-to-leading order term, in our study, allows us to estimate when the series expansion stops converging, and when numerical techniques need to be employed instead.

We find that the effect of $H_{ab}$ is small, while the series expansion remains valid, but grows with time. In particular, we find that while the effect of the $H_{ab}$ on the expansion of LRS curves can be at the level of $1\%$ when the number of masses in the Universe is small (as in \cite{KHB}), but that this number decreases as the number of masses in the universe is increased (when comparing at the same time, measured in terms of $m$). While the effect on the expansion rate is small for larger lattices, however, it does appear that the effect of $H_{ab}$ on the expansion is, in some sense, cumulative. That is, the difference in the curve length that results from including the effect of $H_{ab}$ tends to increase over time, as the magnitude of the magnetic Weyl tensor increases over time. When comparing lattices at the same cosmological time ({\it i.e.} at the same time, as measured in units of $n \times m$), it therefore appears that the effect of $\curl H_{ab}$ on the evolution of the LRS curve increases as the number of masses in the universe is increased. Whether or not similar result holds in cosmological models that expand for all time, rather than re-collapsing, remains to be seen.

It should be noted, however, that while the effects we find are always small (less than $1\%$, in most cases), the Taylor series approximation used to derive them usually breaks down on time scales that are shorter than the age of the universe. This is especially true in the larger lattices, where it can be seen that increasing the number of masses in the universe results in the series approximation breaking down at earlier cosmological times. This result holds for all of the LRS curves that we studied, but seems to be especially true of the cell edges (which, unfortunately, are probably the best indicators of the scale of the cosmology as a whole). To reliably follow the evolution of these curves any further will require more advanced techniques.

It is interesting to see that the effect of higher-order terms in Einstein's equations can lead to non-negligible effects in the large-scale expansion of space. One may note, for example, that it is not until we get to order $t^6$ in the Taylor series expansion in Eq.\ (\ref{da1}) that the influence of $\curl H_{ab}$ becomes non-zero at all. In a weak-field expansion of the gravitational field, about a suitably chosen background, this would probably correspond to quite a high order in perturbations. Nevertheless, the terms involved grow rapidly over cosmological time-scales, until they come close in magnitude to the leading-order terms represented by $E_{ab}$. This is remiscent of the gravitational wave memory effect \cite{memory}, where the small effects of non-linear gravity accumulate over time until (in the case of gravitational waves from astrophysical sources) they are comparable with the magnitude of the linear, leading-order terms.

\begin{acknowledgements}
We are grateful to R.~Tavakol, M.~Korzy\'{n}ski, I.~Hinder \& E.~Bentivegna for helpful conversations and correspondence. TC is supported by the STFC and DG by an AARMS postdoctoral fellowship.
\end{acknowledgements}

\end{document}